\documentclass[11pt]{article}
\usepackage{feynarts}
\usepackage[dvips]{graphicx,color}
\usepackage{amsmath}
\usepackage{amssymb}
\usepackage{cite}
\newcommand{\half}{\frac{1}{2}}
\newcommand{\beq}{\begin{equation}}
\newcommand{\eeq}{\end{equation}}

\newcommand{\beqn}{\begin{eqnarray}}
\newcommand{\eeqn}{\end{eqnarray}}

\newcommand{\nn}{\nonumber}
\newcommand{\idel}{i\delta}
\newcommand{\ud}{\mathrm{d}}
\makeatletter
\DeclareRobustCommand{\Cpp}
{\valign{\vfil\hbox{##}\vfil\cr
   \textsf{C\kern-.1em}\cr
   $\hbox{\fontsize{\sf@size}{0}\textbf{+\kern-0.05em+}}$\cr}%
}
\makeatother

\title{\vspace*{-4.cm}
\begin{flushright}\begin{normalsize}
Cavendish-HEP-07/09\\
Edinburgh 2007/21\end{normalsize}
\end{flushright}
\vspace*{2cm}
Loop induced interference effects in Higgs Boson plus two jet\\ production at the LHC}
\author{J.~R.~Andersen$^{a}$, T.~Binoth$^{b}$, G.~Heinrich$^{b}$,
  J.~M.~Smillie$^{a}$\\\mbox{}\\$^a$Cavendish Lab, JJ Thomson Avenue, Cambridge, CB3
  0HE, UK\\
$^b$School of Physics, The University of Edinburgh, Edinburgh EH9 3JZ, UK}
\begin{document}

\maketitle


\begin{abstract}
  We calculate the order $\mathcal{O}(\alpha^2\alpha_s^3)$
  interference effect between the gluon fusion and weak boson fusion
  processes allowed at the one-loop level in Higgs boson plus 2 jet production
  at the LHC.  
  The corresponding one-loop amplitudes, which have not been considered in the
  literature so far, are evaluated analytically using dimensional regularisation 
  and the necessary master integrals with massive propagators are reported.  
  It is discussed in detail how various mechanisms conspire to make this 
  contribution numerically negligible for experimental studies at the LHC.
\end{abstract}


\section{Introduction}
\label{sec:introduction}

\begin{figure}[tbp]
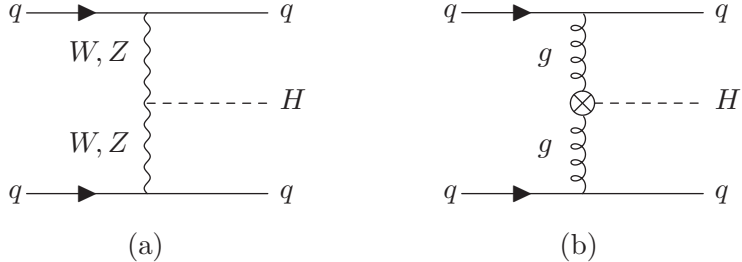

\vspace{-0.5cm}
  \unitlength=1bp
  \hspace{1.4cm}
  \begin{feynartspicture}(100,200)(1,1)
    \FADiagram{}
    \FAProp(10.,10.)(30.,10.)(0.,){/Straight}{0}
    \FALabel(8.5,9.2)[bl]{$q$}
    \FALabel(31.,9.2)[bl]{$q$}
    \FAProp(10.,25.)(30.,25.)(0.,){/Straight}{0}
    \FALabel(8.5,24.2)[bl]{$q$}
    \FALabel(31.,24.2)[bl]{$q$}
    \FAProp(20.,10.)(20.,25.)(0.,){/Sine}{0}
    \FALabel(13.5,20.5)[bl]{$W,Z$}
    \FALabel(13.5,13.)[bl]{$W,Z$}
    \FAProp(20.,17.5)(30.,17.5)(0.,){/ScalarDash}{0}
    \FALabel(31.,17.)[bl]{$H$}
    \FALabel(15.,10.)[cc]{$\blacktriangleright$}
    \FALabel(15.,25.)[cc]{$\blacktriangleright$}
  \end{feynartspicture}
  \hspace{2.cm}
  \begin{feynartspicture}(100,200)(1,1)
    \FADiagram{}
    \FAProp(10.,10.)(30.,10.)(0.,){/Straight}{0}
    \FALabel(8.5,9.2)[bl]{$q$}
    \FALabel(31.,9.2)[bl]{$q$}
    \FAProp(10.,25.)(30.,25.)(0.,){/Straight}{0}
    \FALabel(8.5,24.2)[bl]{$q$}
    \FALabel(31.,24.2)[bl]{$q$}
    \FAProp(20.,10.)(20.,16.5)(0.,){/Cycles}{0}
    \FAProp(20.,18.5)(20.,25.)(0.,){/Cycles}{0}
    \FALabel(16.3,20.5)[bl]{$g$}
    \FALabel(16.3,13.)[bl]{$g$}
    \FAProp(21.,17.5)(30.,17.5)(0.,){/ScalarDash}{0}
    \FALabel(31.,17.)[bl]{$H$}
    \FAProp(19.,17.5)(21.,17.5)(1.,){/Straight}{0}
    \FAProp(19.,17.5)(21.,17.5)(-1.,){/Straight}{0}
    \FAProp(20.7,16.8)(19.3,18.2)(0.,){/Straight}{0}
    \FAProp(19.3,16.8)(20.7,18.2)(0.,){/Straight}{0}
    \FALabel(15.,10.)[cc]{$\blacktriangleright$}
    \FALabel(15.,25.)[cc]{$\blacktriangleright$}
  \end{feynartspicture}  

  \vspace{-3.2cm}
  \hspace{4.5cm} (a) \hspace{5cm} (b)
  \caption{(a) The WBF process for Higgs production in the Standard Model and
    (b) the equivalent gluon-fusion diagram mediated through a top-loop.}
   \vspace{-0.5cm}
   \label{fig:VBF}
\end{figure}

One of the main tasks for the experimental and theoretical programme in connection with
the CERN LHC is to investigate the mechanism of electro-weak symmetry breaking. Central to
this study would be the measurement of the couplings of any observed Higgs scalar to the
electro-weak bosons. This can be performed either by studying the decays $H\to
ZZ,WW$~\cite{Choi:2002jk,Rainwater:1999sd} with contributions from all production
channels, or the production process $pp\to
Hjj$~\cite{Cahn:1983ip,Dicus:1985zg,Altarelli:1987ue} through weak boson fusion
(WBF)~\cite{Plehn:2001nj}, as shown in Fig.~\ref{fig:VBF}(a),  with contributions from all
identifiable decay channels. The Higgs plus two jet signature also receives contributions
from Higgs boson production through gluon-fusion mediated through a top-loop, as
illustrated in Fig.~\ref{fig:VBF}\,(b).
However, the Higgs plus dijet-sample can be biased towards WBF by suppressing
the gluon-fusion channel through a combination of cuts, requiring both
well-separated jets (effectively suppressing the largest component of
gluon-initiated processes) and suppressing events with further central jets
(produced predominantly in the gluon-fusion channel, since here the two
tagging jets are colour-connected).

For the gluon fusion process, the first radiative corrections have been
calculated within QCD~\cite{Campbell:2006xx,DelDuca:2004wt} using the heavy
top mass effective
Lagrangian\cite{Wilczek:1977zn,Dawson:1990zj,Djouadi:1991tk}. For the WBF,
both the radiative corrections within
QCD~\cite{Han:1992hr,Djouadi:1991tk,Figy:2003nv,Berger:2004pc} and the
electro-weak sector~\cite{Ciccolini:2007jr} have been calculated. The
radiative corrections to the WBF channel are small, $3 \% - 6\% 
$, and there is
even partial numerical cancellation between the QCD and electro-weak
contributions.  It would therefore seem that the Higgs coupling to
electro-weak bosons can be very cleanly studied with a $Hjj$-sample.

Until recently, the irreducible contamination in the extraction of the
$ZZH$-coupling from interference between the gluon fusion and WBF processes
was ignored in the literature. At tree level, such interference is only
allowed in amplitudes where the two quarks have the same  flavours, but their 
 contribution is
kinematically suppressed by the requirement
of a $t\leftrightarrow u$-channel crossing, as discussed in
Ref.\cite{Andersen:2006ag}. These interference terms were  later also 
included in the calculation reported in Ref.\cite{Ciccolini:2007jr}, 
where the full 
electroweak corrections have been calculated, 
and which also took into
account other crossing-suppressed one-loop amplitudes. 

In the present paper we will report on the calculation of the processes allowed at the
one-loop level which do not suffer from the suppression stemming from the requirement of
a $t\leftrightarrow u$-crossing.  As will be explained below, one finds at order
$\mathcal{O}(\alpha^2\alpha_s^3)$ an interference term between the gluon- and $Z$-induced
amplitude which is not allowed at $\mathcal{O}(\alpha^2\alpha_s^2)$ by colour
conservation.  
The $W$-induced amplitudes are crossing-suppressed and therefore not taken into account.   
The diagrams where the vector boson is in the s-channel can be safely neglected 
because they are strongly suppressed by the  WBF cuts.

Given that electroweak corrections to the WBF process, which are formally an order
$\mathcal{O}(\alpha)$ correction to an $\mathcal{O}(\alpha^4)$ process, 
have been shown to be relevant for this important process
\cite{Ciccolini:2007jr}, simple power counting alone suggests that the size of the
irreducible contamination due to the discussed interference effect should be checked. We
will elaborate below that arguments in the literature which are based on simplified
assumptions do not capture all effects found by doing the full one-loop calculation.

In the following section we will briefly sketch the calculation before discussing our
results in section \ref{sec:results}, which are summarized in the conclusions.  The
appendix contains an extensive  list of  the master integrals needed for this calculation.
Most of these integrals have not been reported in the literature so far.

\section{The Calculation}
\label{sec:calculation}
\begin{figure}[tbp]
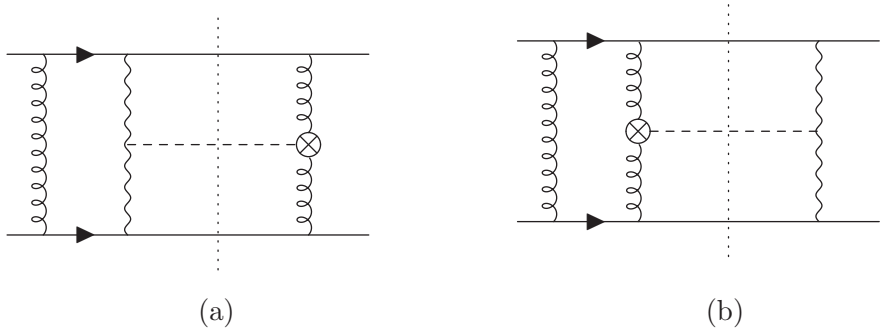


\vspace{-0.5cm}
  \unitlength=1bp
  \hspace{1.4cm}
  \begin{feynartspicture}(100,190)(1,1)
    \FADiagram{}
    \FAProp(0.,10.)(30.,10.)(0.,){/Straight}{0}
    \FAProp(0.,25.)(30.,25.)(0.,){/Straight}{0}
    \FAProp(10.,10.)(10.,25.)(0.,){/Sine}{0}
    \FAProp(25.,10.)(25.,16.5)(0.,){/Cycles}{0}
    \FAProp(25.,18.5)(25.,25.)(0.,){/Cycles}{0}
    \FAProp(10.,17.5)(24.,17.5)(0.,){/ScalarDash}{0}
    \FAProp(17.5,28.)(17.5,7.)(0.,){/GhostDash}{0}
    \FAProp(3.,10.)(3.,25.)(0.,){/Cycles}{0}
    \FALabel(6.5,10.)[cc]{$\blacktriangleright$}
    \FALabel(6.5,25.)[cc]{$\blacktriangleright$}

    \FAProp(24.,17.5)(26.,17.5)(1.,){/Straight}{0}
    \FAProp(24.,17.5)(26.,17.5)(-1.,){/Straight}{0}
    \FAProp(25.7,16.8)(24.3,18.2)(0.,){/Straight}{0}
    \FAProp(24.3,16.8)(25.7,18.2)(0.,){/Straight}{0}
  \end{feynartspicture}
\hspace{3cm}
  \begin{feynartspicture}(100,200)(1,1)
    \FADiagram{}
    \FAProp(0.,10.)(30.,10.)(0.,){/Straight}{0}
    \FAProp(0.,25.)(30.,25.)(0.,){/Straight}{0}
    \FAProp(10.,10.)(10.,16.5)(0.,){/Cycles}{0}
    \FAProp(10.,18.5)(10.,25.)(0.,){/Cycles}{0}
    \FAProp(25.,10.)(25.,25.)(0.,){/Sine}{0}
    \FAProp(11.,17.5)(25.,17.5)(0.,){/ScalarDash}{0}
    \FAProp(17.5,28.)(17.5,7.)(0.,){/GhostDash}{0}
    \FAProp(3.,10.)(3.,25.)(0.,){/Cycles}{0}

    \FAProp(9.,17.5)(11.,17.5)(1.,){/Straight}{0}
    \FAProp(9.,17.5)(11.,17.5)(-1.,){/Straight}{0}
    \FAProp(10.7,16.8)(9.3,18.2)(0.,){/Straight}{0}
    \FAProp(9.3,16.8)(10.7,18.2)(0.,){/Straight}{0}

    \FALabel(6.5,10.)[cc]{$\blacktriangleright$}
    \FALabel(6.5,25.)[cc]{$\blacktriangleright$}
  \end{feynartspicture}

\vspace{-2.7cm}
\hspace{4.1cm} (a) \hspace{6.cm} (b)
\caption{\label{fig:interfdiags} Example of contributing one-loop interference terms: (a)
  $\mathcal{M}_{gZ}\mathcal{M}_g^*$ and (b) $\mathcal{M}_{gg}\mathcal{M}_Z^*$. There are
  four contributing topologies for each gluon-fusion and $Z$-fusion process. }
\end{figure}

\begin{figure}[tbp]
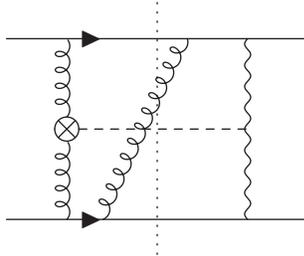

\centering
\vspace{-0.5cm}
  \unitlength=1bp
  \hspace{-3cm}
  \begin{feynartspicture}(100,180)(1,1)
    \FADiagram{}
    \FAProp(5.,10.)(30.,10.)(0.,){/Straight}{0}
    \FAProp(13.,10.)(20.,25.)(0.,){/Cycles}{0}
    \FAProp(5.,25.)(30.,25.)(0.,){/Straight}{0}
    \FAProp(10.,10.)(10.,16.5)(0.,){/Cycles}{0}
    \FAProp(10.,18.5)(10.,25.)(0.,){/Cycles}{0}
    \FAProp(25.,10.)(25.,25.)(0.,){/Sine}{0}
    \FAProp(11.,17.5)(25.,17.5)(0.,){/ScalarDash}{0}
    \FAProp(17.5,28.)(17.5,7.)(0.,){/GhostDash}{0}

    \FAProp(9.,17.5)(11.,17.5)(1.,){/Straight}{0}
    \FAProp(9.,17.5)(11.,17.5)(-1.,){/Straight}{0}
    \FAProp(10.7,16.8)(9.3,18.2)(0.,){/Straight}{0}
    \FAProp(9.3,16.8)(10.7,18.2)(0.,){/Straight}{0}

    \FALabel(12,10.)[cc]{$\blacktriangleright$}
    \FALabel(12,25.)[cc]{$\blacktriangleright$}
  \end{feynartspicture}  

  \vspace{-3.2cm}
  \caption{One of the real emission processes which contributes at the 
matrix element
    squared level.}
  \label{fig:realsquared}
\end{figure}

We set out to calculate the helicity amplitudes necessary to form the loop
interference terms  and the real emission
contributions. Sample diagrams are shown in 
Figs.~\ref{fig:interfdiags} and \ref{fig:realsquared}. As discussed in
Ref.\cite{Andersen:2006ag} this is the lowest order contribution to the
interference between $ZZH$ and $ggH$-processes for non-identical quark
flavours and helicity configurations, and for identical quark and helicity
configurations the loop amplitudes are the first order which does not require
a kinematically disfavoured crossing.

The amplitudes have four non-zero helicity components, which we label by $++++$, $----$,
$-++-$ and $+--+$. Due to parity invariance of the kinematical part of the amplitudes,
only two of them are independent. By using the spinor helicity formalism we have defined
projection operators on each of these amplitudes.  In practice, we calculate the
amplitudes with all momenta incoming (and therefore summing to zero); to map to physical
scattering kinematics, crossing relations are applied easily in the end.

The leading order amplitudes, denoted by $\mathcal{M}_Z$ 
and $\mathcal{M}_g$ (see Fig.~\ref{fig:interfdiags}), are proportional to a colour singlet
and a colour octet term.  The colour singlet is formally of order
$\mathcal{O}({\alpha}^2)$ whereas the octet is of order
$\mathcal{O}(\alpha_s^2)$.  The virtual corrections, which we call
$\mathcal{M}_{gZ}$ and $\mathcal{M}_{gg}$ respectively, are mixtures of octet and singlet
terms.  For the intereference term we need to consider only the octet part of
$\mathcal{M}_{gZ}$ and the singlet part of $\mathcal{M}_{gg}$.  One finds that only four
one-loop five-point topologies for each amplitude survive this colour projection. As was
already pointed out in Ref.\cite{Andersen:2006ag}, the colour singlet cannot interfere
with the colour octet tree amplitude for different quark flavours. However, a new colour
channel opens up at order $\mathcal{O}(\alpha^2\alpha_s^3)$ which is neither flavour nor
kinematically suppressed.

The loop amplitudes require the evaluation of one-loop five-point tensor
integrals with partly massive propagators and external legs. We apply the reduction
algorithm outlined in Ref.~\cite{Binoth:1999sp,Binoth:2005ff} to express each
Feynman diagram as a linear combination of 1-, 2-, and 3-point functions in
$D=4-2\epsilon$ dimensions and 4-point functions in $D$=6. The same 
algorithm has been successfully applied to a number of one-loop
computations and further details can be found elsewhere 
\cite{Binoth:2006mf,Binoth:2006ym,Binoth:2005ua,Binoth:2003xk}.
The coefficient
of each integral, which is a rational polynomial in terms of Mandelstam
variables $s_{ij}=(p_i+p_j)^2$ and masses, was evaluated symbolically using
FORM~\cite{Vermaseren:2000nd} and simplified using
\textsc{Mathematica}~\cite{wolfram05:mathem}.  
Both steps were fully automated. 
The algebraic expressions were checked by independent implementations, both
amongst the authors and with another group~\cite{Dixon:priv}.

After the algebraic reduction, all helicity amplitudes for both cases, gluon
and weak boson fusion, were obtained as linear combinations of a certain
number of scalar integrals. We choose this basis of so-called master integrals (MIs) in
accordance with Ref.~\cite{Binoth:2007ca}, i.e. our MI's are $D$-dimensional two-point and
three-point functions $(I^D_2,I^D_3)$, and ($D$+2)-dimensional four-point functions 
$(I_4^{D+2})$.  Schematically 
\beqn 
\mathcal{M} =
\sum\limits_{j,\alpha} k_{j\alpha} I_j(\{s_\alpha,m_\alpha\}) \; , \quad I_j
\in \{I_2^D,I_3^D,I_4^{D+2} \} 
\eeqn 
where the summation over $\alpha$ indicates the summation over different argument lists
$\{s_\alpha,m_\alpha\}$ of the relevant MI.  The conventions for the arguments and the
analytic forms of these are given in the appendix.  No one-point functions appear in the
reduction, and also two-point functions are absent in the amplitudes of
$\mathcal{M}_{gZ}$. Furthermore, coefficients of some of the integrals which arise in
several topologies sum to zero:  if the tree resulting from a cut of an internal line 
of a master integral
corresponds to helicity forbidden tree level processes one can immediately infer the
vanishing of the corresponding coefficient. In our algebraic tensor reduction approach we
are able to verify such cancellations and enforce them analytically before the numerical
evaluation of the cross section (for a nontrivial example see Ref.~\cite{Binoth:2007ca}).

The coefficients we obtain through this procedure can be too large to be of use in a
printed form (their simple polynomial structure means that numerical evaluation is,
however, fast). We performed several algebraic checks of relations between coefficients of
different topologies and helicity configurations.  The coefficients are included as
supplementary material to this paper.

As most of the required integrals are not provided in the literature, we have
evaluated representations in terms of analytic functions valid in all kinematic
regions.  We give our result for these integrals in the appendix, as they
might be of use for other calculations\footnote{Of course all 
  finite integrals can in principle be evaluated by using the LoopTools
  package~\cite{Hahn:1998yk}. The $D=6$ boxes can be written as linear
  combinations of 3- and 4-point functions in D=4. If IR divergences are
  present, a small regulator mass must be used, but its dependence
  can be made arbitrarily small as the D=6 box integral is IR finite.}.  
  
The IR structure of the interference term is very simple. It
is easily extracted from the
result by focusing on the IR divergent triangles. All MIs with single poles drop out. 
Only triangles with double poles survive. This results in  an expression 
$$\sim \alpha_s C_F [ (-s_{13})^{-\epsilon} +
(-s_{24})^{-\epsilon} - (-s_{12})^{-\epsilon} -(-s_{34})^{-\epsilon} ]/\epsilon^2$$ 
in which all double poles cancel and only  sub-leading soft
divergences survive.

The virtual corrections to the interference term have to be combined with the real
emission part shown in Fig.~\ref{fig:realsquared}.  In accordance with the virtual corrections, 
the collinear IR divergences from the three-parton final states integrate to zero, leaving only a soft
divergence proportional to $1/\varepsilon$ in dimensional regularisation.  Due to the
simple structure of the divergences, we have used the phase space slicing
method\cite{Giele:1991vf,Giele:1993dj}
to isolate the  IR divergences from the real radiation part.
We have checked that
the remaining single poles cancel exactly when combining the real
emission with the virtual part.  The phase space integration and the numerical evaluation of
integrals and coefficients is coded in a \Cpp\, program allowing for a
flexible implementation of cuts and observables.

\section{Results}
\label{sec:results}

This study aims at investigating a possible pollution of the clean
extraction of the $ZZH$ vertex structure by the interference
terms. 
Therefore we will apply the  cuts summarised in Table \ref{tab:cuts},
which are generally used for the selection of WBF
events~\cite{DelDuca:2001fn} over the gluon fusion ``background''.
\begin{table}[tbp]
  \centering
  \begin{tabular}{|rl||rl|}
    \hline
    $p_{a_T}$, $p_{b_T}$ & $> 20$ GeV & $\eta_a\cdot\eta_b$& $<0$ \\
    $\eta_j$ & $<$ 5 & $\vert \eta_a-\eta_b \vert$ & $> 4.2$ \\
    $s_{ab}$ & $>$ (600 GeV)$^2$ & & \\ \hline
  \end{tabular}
  \caption{The cuts used in the following analysis which bias the Higgs Boson
    plus dijet sample towards WBF.  The suffices $a,b$ label the tagged jets.}
  \label{tab:cuts}
\end{table}
Our input parameters for the numerical studies are taken from either the parton density
function-fit\cite{Martin:2004ir} in the case of $\alpha_S(M_Z^2)$ and the Review of Particle
Physics\cite{PDBook} for the others.
\begin{align}
  \begin{split}
    &\alpha_s(M_Z^2) = 0.1205\, ,\qquad g^2=\frac{G_F}{\sqrt 2} \frac 1 {8
      M_W^2} \, ,\qquad G_F = \,1.16637\times
    10^{-5}~\mathrm{GeV}^{-2}\\& M_Z = 91.1876~\mathrm{GeV}\,
    ,\quad M_W = 80.425~\mathrm{GeV}\, ,\quad\sin^2\theta_W = 0.2312\;.
  \end{split}
\end{align}
We have checked that variations of the numerical values chosen for the WBF
cuts have no impact on the conclusions, nor does the exact value of the Higgs
boson mass, which we set to 115 GeV unless otherwise stated.  The same is true
for the choice of parton sets; we choose to use the NLO set from
Ref.\cite{Martin:2004ir}, and use 2-loop running for $\alpha_s$, in accordance with the 
chosen pdfs.

We observe that in all the flavour and helicity
channels, the finite contribution from the 3-parton final state is numerically
negligible.  In fact, in the case at hand, the only r\^ole of this real
emission is to cancel the divergences which arise from the one-loop diagrams.

As an interference effect proportional to $2
\mathrm{Re}(\mathcal{M}_{gg}\mathcal{M}_Z^*+\mathcal{M}_{gZ}\mathcal{M}_g^*)$,
the result is not necessarily positive definite.  In fact, the sign of the
interference contribution depends on the azimuthal angle between the two
tagging jets, $\Delta\phi_{jj}$. Because of the event topology with two well
separated jets, it becomes possible to define an orientation of the azimuthal
angle which allows observability in the whole range of $[-\pi,\pi)$, as
pioneered in Ref.\cite{Plehn:2001nj,Hankele:2006ma}.  $\Delta\phi_{jj}$ is then defined
through
\begin{align}
  \begin{split}
    \label{eq:phijj}
  {|p_{+_T}||p_{-_T}|} \cos \Delta\phi_{jj}&= p_{+_T}\cdot
      p_{-_T},\\
    2 |p_{+_T}|| p_{-_T}| \sin \Delta\phi_{jj}&=\varepsilon_{\mu\nu\rho\sigma}
    b_+^\mu p_+^\nu
    b_-^\rho p_-^\sigma,
  \end{split}
\end{align}
where $b_+$ ($b_-$) are unit vectors in positive (negative) beam direction, 
and likewise for the jet momenta $p_\pm$. The cuts
ensure that the two tagging jets lie in opposite hemispheres.

\begin{figure}[tbp]
  \centering
  \includegraphics[width=0.48\textwidth]{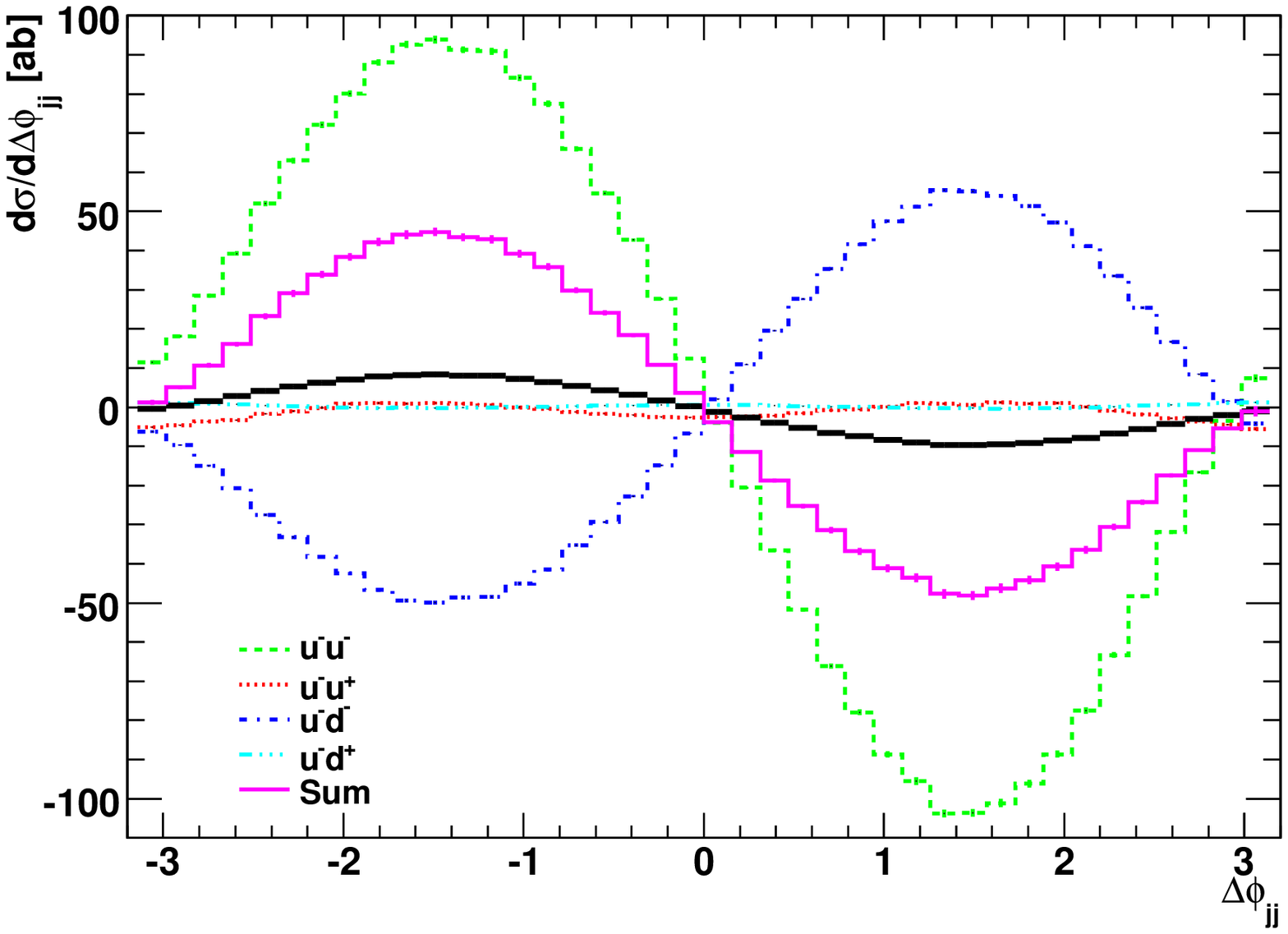}
  \includegraphics[width=0.48\textwidth]{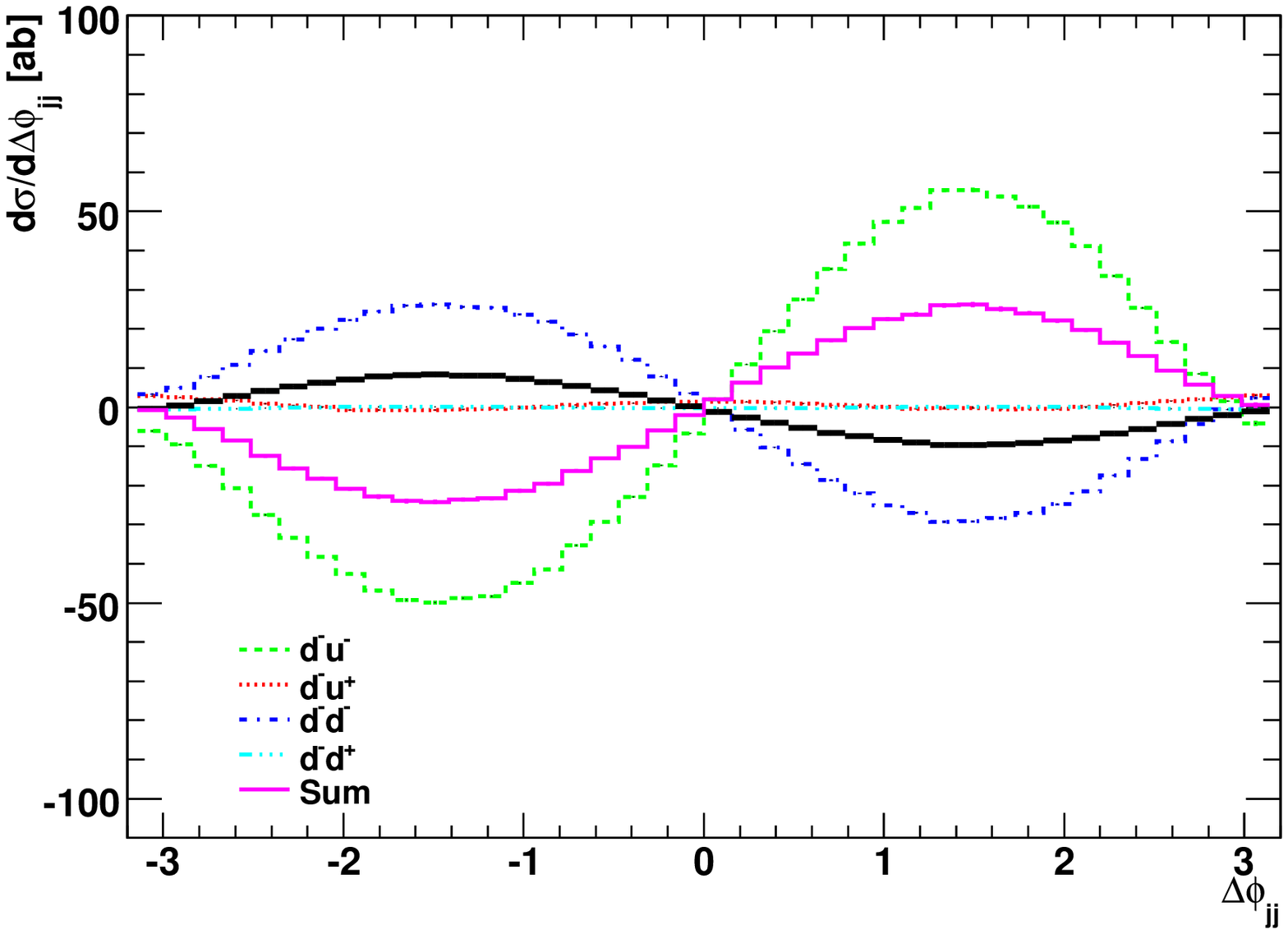}
  \caption{The $\Delta\phi_{jj}$-distribution for different  helicity-configurations of the 
    valence quarks only. 
    The purple histogram labelled ``Sum'' indicates
    the sum over the four contributions shown. The sum over \emph{all}
    valence quark flavour and helicity assignments is shown in the black histogram.}
  \label{fig:flavhel_val}
\end{figure}

\begin{figure}[tbp]
  \centering
  \includegraphics[width=0.48\textwidth]{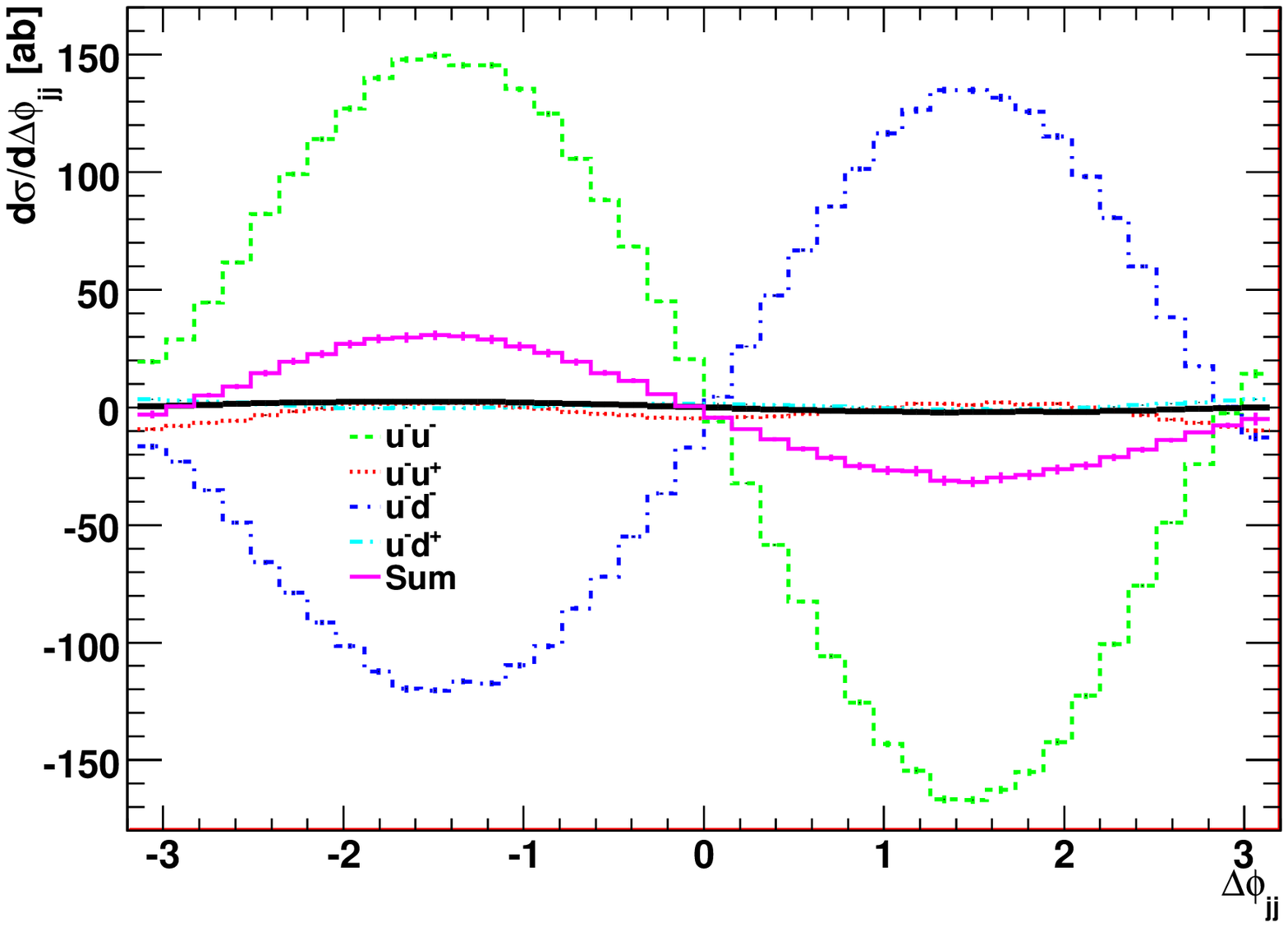}
  \includegraphics[width=0.48\textwidth]{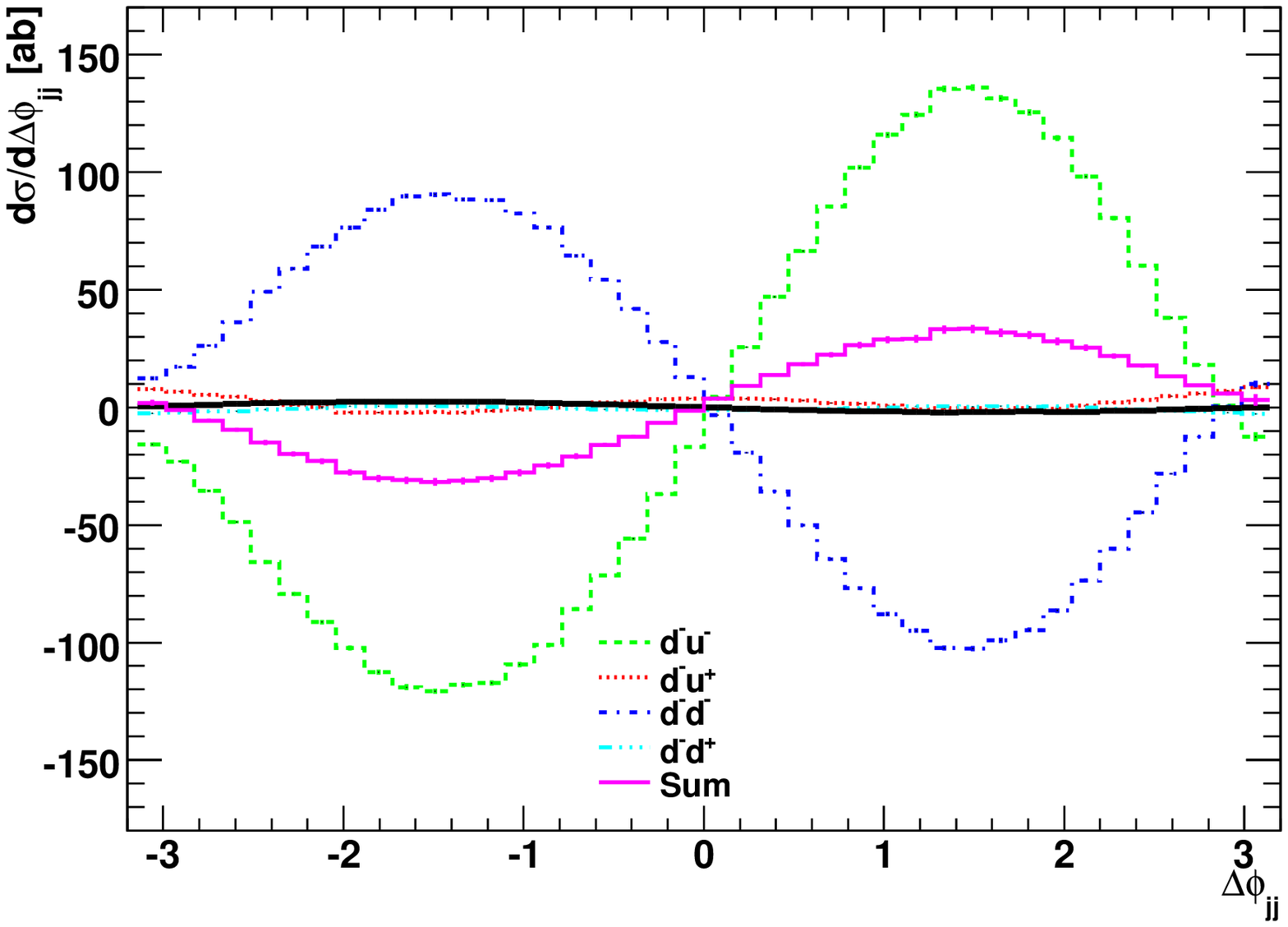}
  \caption{The $\Delta\phi_{jj}$-distribution for various flavour and
    helicity-configurations. The purple histogram labelled ``Sum'' indicates
    the sum over the four contributions shown. The sum over \emph{all}
    flavour and helicity assignments including all sea flavours is shown in the 
    black histogram.}
  \label{fig:flavhel_seaval}
\end{figure}

Figure~\ref{fig:flavhel_val} displays the contribution to the distribution in
$\Delta\phi_{jj}$ from the interference terms for various helicity and flavour
configurations of the valence quarks only, for a Higgs boson mass of 115~GeV.
Figure~\ref{fig:flavhel_seaval} displays the contribution to the distribution in
$\Delta\phi_{jj}$ including sea quarks.

Due to the oscillatory behaviour, the total integrated cross section does not at all tell
the full story about the size of the impact on the $\Delta\phi_{jj}$-distribution; for
example, the integral of the contribution from 
the sea and valence up-quarks with negative helicity, denoted by $u^{-}u^{-}$, 
 is $+5$~ab, while the
distribution peaks at more than $150$~ab/rad. 
If only valence quarks are considered, the integral is $-30$~ab, while the
distribution peaks at $\sim 90$~ab/rad. 

There is an accidental cancellation of sea and valence
quark contributions which leads to the fact that
the sum over all flavour and helicity assignments peaks at around $2$~ab/rad only, 
with an integrated effect of $1.19\pm0.07$~ab, where the error is due to the 
numerical integration.
Note that the integral of the absolute
value of the $\phi_{jj}$ distribution, 
$$\int_{-\pi}^\pi d \Delta\phi_{jj} |\frac{d\sigma}{d\Delta\phi_{jj}}|\;\;,$$
is a useful  measure of the impact of the
interference effect on the extraction of the $ZZH$-vertex. 
This integral evaluates  to $9.1\pm 0.1$~ab, an order of magnitude larger.
The total integral over the the absolute value of the fully differential 
cross section leads  to $29.59\pm 0.07$~ab.

\begin{figure}[htb]
  \centering
  \includegraphics[width=0.6\textwidth]{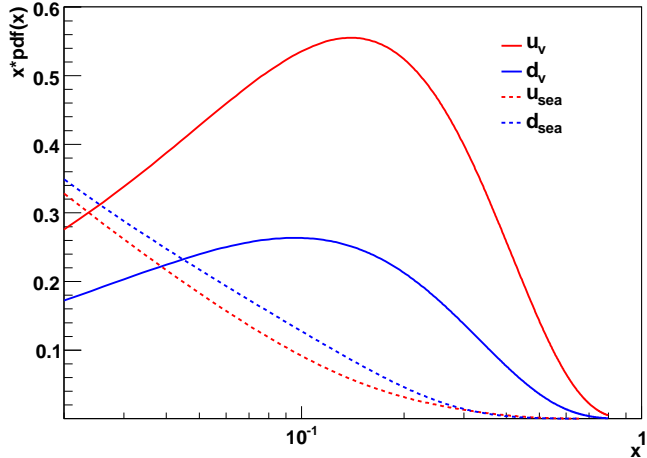}
  \caption{The required parton distribution functions\,\cite{Martin:2004ir} 
  in the relevant region of Bj\"orken
    $x$ for $Q^2=400\ \mathrm{GeV}^2$.}
  \label{fig:pdfs}
\end{figure}

\begin{figure}[htb]
  \centering
  \includegraphics[width=0.48\textwidth]{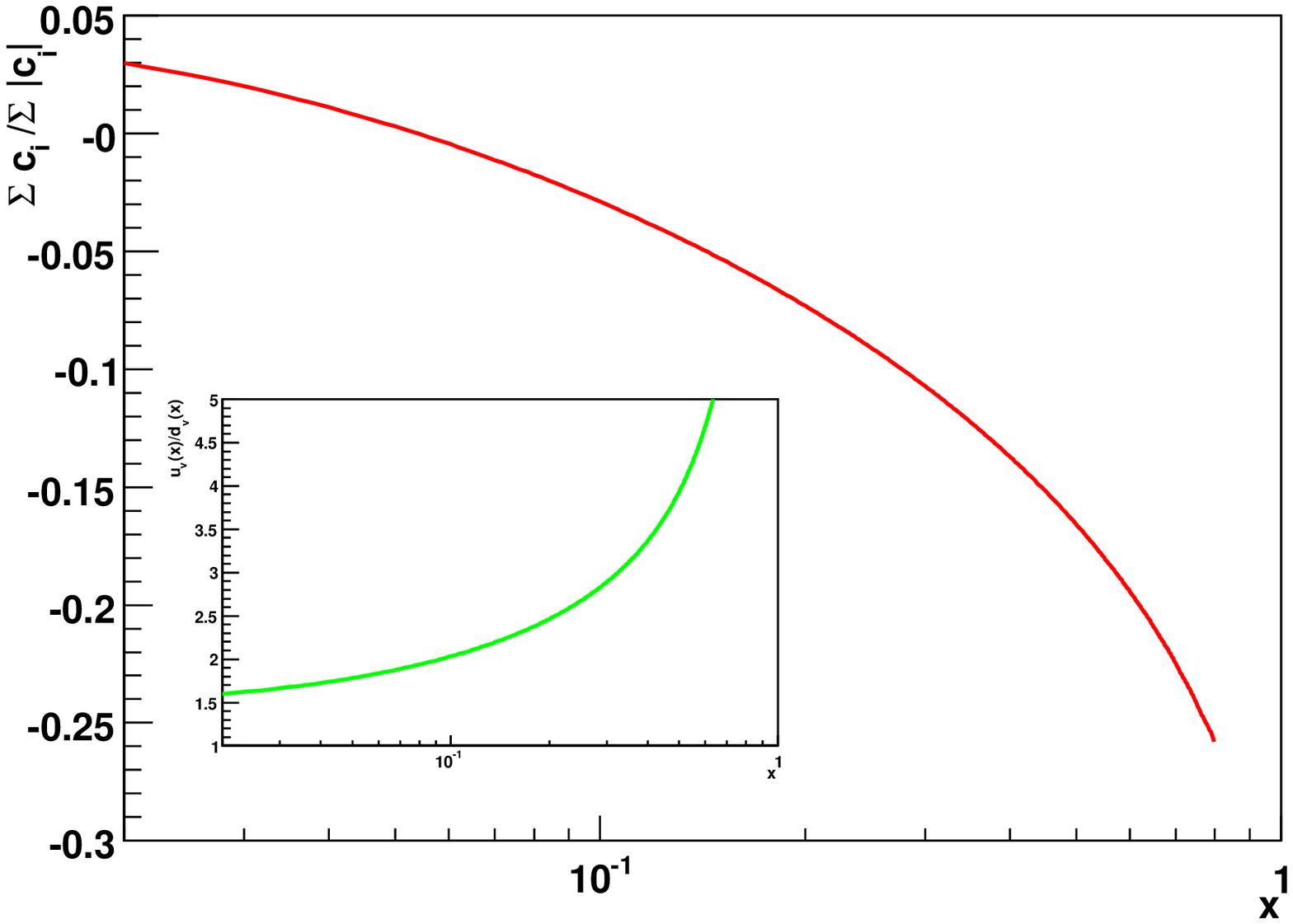}
  \includegraphics[width=0.48\textwidth]{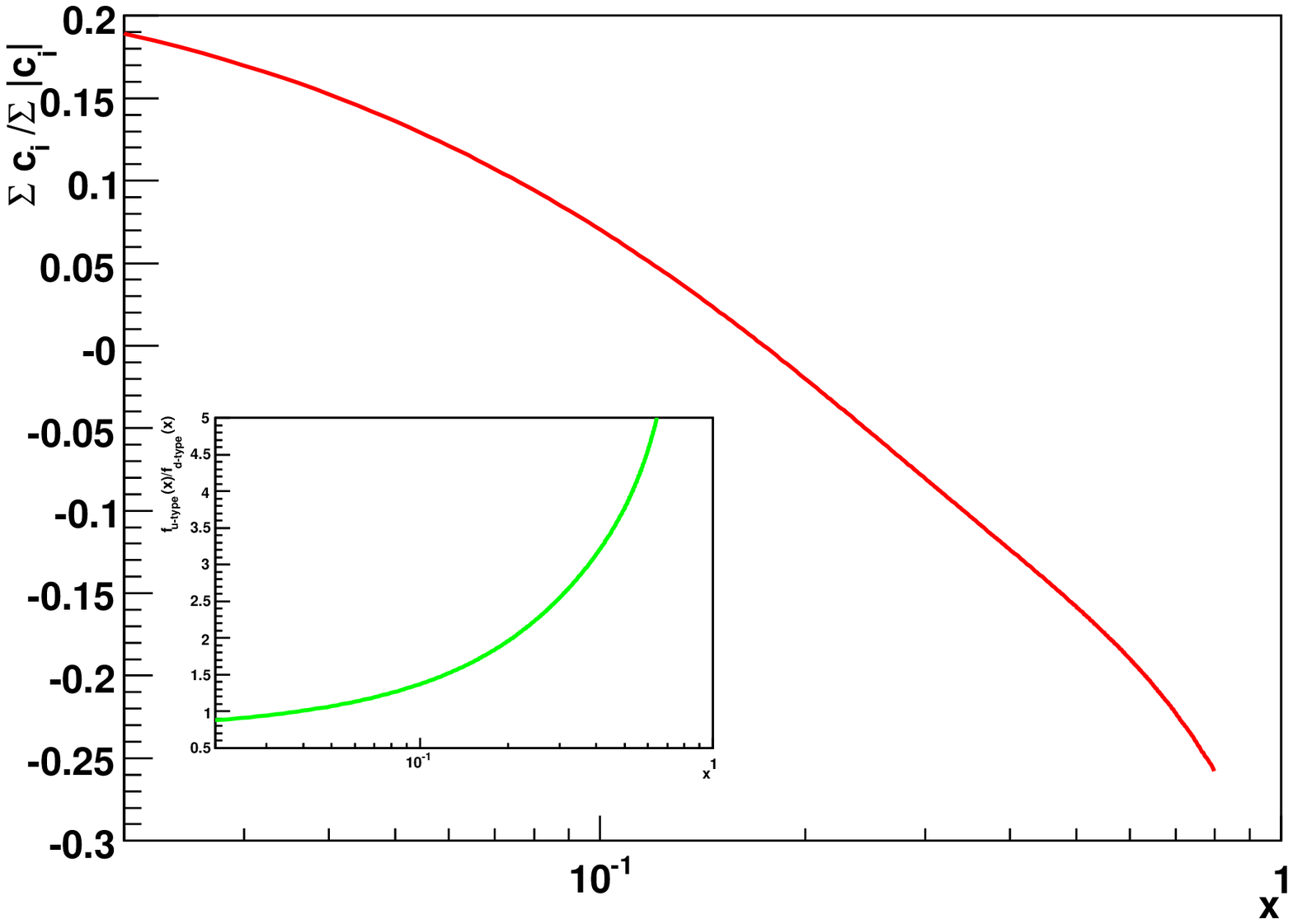}
  \caption{The pdf-weighted sum of the $SU(2)\times U(1)$-charges divided by the
    sum of the absolute weighted charges for valence quarks only (left)
    and valence and sea quarks (right). The left figure illustrates how the apparent
    almost complete cancellation expected from the Quark Model is in fact not
    as severe once the pdfs are taken into account, as discussed in the
    text. The insert shows the ratio of the valence $u$-quark distribution to
    the valence $d$-quark distribution for the pdf set MRST 2004 (NLO)\cite{Martin:2004ir}.
    The right plot shows that the inclusion of sea quarks does actually alter the pdf-weighted
    sum in the relevant $x$ range such that there are additional compensations when integrating 
    over $x$.}
  \label{fig:pdfratios}
\end{figure}

As can be readily seen, there is a cancellation between the contribution from
each flavour and helicity assignment; this is because the sign of quark
couplings to the $Z$-boson becomes relevant, since it is not squared for the
interference. The flavour- and helicity sum for each quark line therefore
leads to some cancellation, resulting from the weak charges listed in
Table~\ref{tab:weakcharges}.
\begin{table}[tbp]
  \centering
  \begin{tabular}{l|rr}
    $q_\lambda$ & $a_{f,\lambda}$&$b_{f,\lambda}$\\\hline
    $u_L$ & 1/2&-2/3\\
    $u_R$ & 0&-2/3\\
    $d_L$ & -1/2&1/3\\
    $d_R$ & 0&1/3\\
  \end{tabular}
  \caption{$SU(2)\times U(1)$ charges for the $Zq\bar q$-couplings, 
    $[a_{f,\lambda}+b_{f,\lambda}\ \sin^2(\theta_{\scriptscriptstyle W})]$}
  \label{tab:weakcharges}
\end{table}
This cancellation was discussed in Ref.~\cite{Forshaw:2007vb} for the valence content
where it was pointed out that if $\sin^2\theta_W=1/4$ and the $u$-valence content of the
proton was exactly twice the $d$-valence content for all Bj\"orken $x$, the
contributions from $uu$, $ud$ and $dd$ would sum to zero.  However, neither
of these approximations is exactly true: $\sin^2\theta_W=0.2312$ and the
parton distribution functions are shown in Fig.~\ref{fig:pdfs}. The
cancellation stemming from the $SU(2)\times U(1)$-flavour and helicity sums
of the valence quarks can be effectively studied by calculating the ratio 
\begin{align}\label{eq:cs}
  \begin{split}
    & \sum_{f,\lambda}
c_{f,\lambda}(x,Q^2)/\sum |c_{f,\lambda}(x,Q^2)|  \qquad \mbox{where}\\
    & c_{f,\lambda}(x,Q^2)= 
  f(x,Q^2)\ [a_{f,\lambda}+b_{f,\lambda}\,\sin^2(\theta_{\scriptscriptstyle W})],
  \end{split}
\end{align}
where $f(x,Q^2)$ is the relevant parton distribution function. This ratio is
plotted in Fig.~\ref{fig:pdfratios} (left). This figure also shows the ratio of the
$u$-valence to the $d$-valence pdf, which influences the cancellation. It
follows that in QCD the cancellation due to $SU(2)\times U(1)$-flavour and
helicity traces amounts to roughly $10^{-1}$ in the most relevant regions of
the pdfs\footnote{We were not able to reproduce the one order of magnitude larger
  suppression factor reported in Ref.~\cite{Forshaw:2007vb} within the naive Quark
  Model.}. If one includes the sea quarks in this argument,
  one sees that the ratio Eq.~(\ref{eq:cs}) is modified significantly, see 
  Fig.~\ref{fig:pdfratios} (right).  Its change of sign in the relavant x-region
  leads to a further reduction of the interference term after integration.

Using the same cuts and value for the mass of the Higgs boson as in the
present study, we have checked that the total contribution to the
$\Delta\phi_{jj}$-distribution from the leading order WBF process (both $Z$
and $W^{+/-}$ included) is relatively flat at around 240~fb/rad. Therefore,
the result of the interference effect reported here is unlikely to be
measurable.


The smallness of the overall effect is in fact also a result of the complex phases 
arising from the full one-loop calculation of the amplitudes. To illustrate this,
we calculate the average value of
\begin{align}
  \frac{|\mathrm{Re}\left(\mathcal{M}_{gZ}\mathcal{M}_g^*+\mathcal{M}_{gg}\mathcal{M}_Z^*\right)|}
  {|\mathcal{M}_{gZ}\mathcal{M}_g^*|+|\mathcal{M}_{gg}\mathcal{M}_Z^*|}.
\end{align}
The average over phase space for this quantity is roughly $20\%$, which illustrates
that the relevant products and sums for the interference effect project out
only a small component of the full complex loop amplitudes.

We have checked that
none of the sources of suppression discussed above depends severely on the mass of the
Higgs boson; since the amplitudes themselves depend on this parameter 
only weakly, the effect of increasing the Higgs boson mass is 
basically nothing but a reduction
of available phase space. This reduction however is very small, since the
partonic centre of mass energy is dominated by the contribution from the jets
rather than the Higgs boson. The exact numerical value chosen for the cuts
also does not affect the relative importance of the interference effect ---
the effect on the interference and the WBF signal is similar, so the relative
importance of the interference is largely unchanged. 

For completeness we list in Table~\ref{tab:interferencevsmh} the
integral of the absolute $\Delta\phi_{jj}$-distribution for various choices
of the renormalisation and factorisation scales. 
\begin{table}[tbp]
  \centering
  \begin{tabular}{c|c|c}
    $\mu_{f,a}=\mu_{r,a}$ &$\mu_{f,b}=\mu_{r,b}$&Integral of $|d\sigma/d\Delta\phi_{jj}|$[ab]\\\hline
    $p_{a_T}$ & $p_{b_T}$ & $9.1\pm0.1$\\
    $m_H/2$ & $m_H/2$& $13.9\pm0.6$\\
    $m_H$ & $m_H$& $9.2\pm0.4$\\
    $2m_H$ & $2m_H$& $6.3\pm0.3$\\
  \end{tabular}
  \caption{The dependence of the interference effect on the choices for
    factorisation and renormalisation scales, with a fixed mass of the Higgs
    boson of 115~GeV. The variation is slightly larger than what would be
    expected from renormalisation scale variations only, since the variations
    in factorisation scale impacts the cancellations between $SU(2)\times
    U(1)$-charges by slightly altering the ratio defined above Eq.~(\ref{eq:cs}).}
  \label{tab:interferencevsmh}
\end{table}

We chose the factorisation and renormalisation scales as in accordance with the
natural scales in the relevant high energy limit (as in
Ref.\cite{DelDuca:2006hk}), i.e. the factorisation scales are set equal to
the transverse momenta of the relevant jet, and the renormalisation scale for
the strong couplings are chosen correspondingly, i.e. one $\alpha_s$
evaluated at each value of the transverse momentum of the jets, and one at
the Higgs mass. However, varying these has no impact on the
conclusions. 


\section{Conclusions}
\label{sec:conclusions}
We have presented the calculation of the loop-induced  
$\mathcal{O}(\alpha^2\alpha_s^3)$ interference effect between the gluon 
fusion and weak boson fusion processes
in Higgs boson plus two jet production at the LHC.

In the context of the weak boson fusion cuts we have evaluated all relevant
one-loop diagrams algebraically and have obtained an analytic
representation of the interference term as a linear combination
of scalar one-loop integrals. The analytic result for all 
the necessary integrals   is presented for general kinematics such
that it can be used in other computations.

Our expressions have been coded into a flexible computer program to test
speculations in the literature about the size of this interference 
contribution. We do confirm by explicit calculation that this 
contribution is too small  to contaminate the 
extraction of the $ZZH$-coupling from WBF processes. 
Interestingly the effect which survives comes dominantly
from the virtual corrections.  
We have analysed in detail why this contribution is so small, 
and instead of a single effect we rather find a conspiracy of 
several mechanisms which can only be completely assessed having the 
full NLO calculation at hand. 

\newpage

The mechanisms basically are
\begin{itemize}
\item accidental cancellations between the sea quark 
      and valence quark contributions 
\item compensations between different weak isospin flavours of the valence quark 
      contributions due to their $SU(2)\times U(1)$ couplings
      in combination with their weights from the (valence) quark content of the proton 
\item cancellations due to destructive interference of the phases from the 
      different contributions.
\end{itemize}
The exact impact of these partly accidental effects has until now not been 
quantified thoroughly and was very hard to assess without 
an explicit calculation. 

As a final  comment we would  like to point  out 
that anomalous couplings which  affect the phases could change the interference 
pattern substantially. However,  the first two cancellation mechanisms 
still being present, the overall contribution is still  expected to be undetectably  
small.

\section*{Acknowledgements}
We would like to thank Lance Dixon for enlightening and encouraging
discussions and important comments. The authors were all supported by the UK Science and Technology
Facilities Council. In addition, the work of TB was supported by the Deutsche
Forschungsgemeinschaft (DFG) under contract number BI 1050/2 and the Scottish
Universities Physics Alliance (SUPA).

\newpage

\begin{appendix}

\renewcommand{\theequation}{\Alph{section}.\arabic{equation}}
\setcounter{equation}{0}
\section{Analytic Results for Master Integrals}
\label{sec:master-integrals}

The appendix contains the Master Integrals (MIs) which occur 
in the reduction of the one-loop pentagon diagrams encountered
in the given calculation.  
The analytic results of the integrals containing only massless propagators have appeared
in the literature, see e.g. \cite{Bern:1992em,Lu:1992ny,Bern:1993kr,Binoth:1999sp}.  Those with
massive propagators have been calculated for this project. 
All finite integrals can also be calculated by the {\tt LoopTools} package \cite{Hahn:1998yk}, 
based on 
\cite{'tHooft:1978xw,vanOldenborgh:1989wn,Denner:1991qq} which we used for checking purposes.

The conventions we use for the scalar triangles and boxes listed below 
are different from the ones 
defined in \cite{Bern:1992em,Bern:1993kr,Binoth:1999sp,Binoth:2005ff}, as we follow 
the LoopTools conventions for the argument lists, 
in  order to comply with the ``Les Houches Accord on Master Integrals". 

\begin{figure}[ht]
\unitlength=1mm
\begin{picture}(140,50)
\put(55,5){\includegraphics[width=4cm, height=4cm]{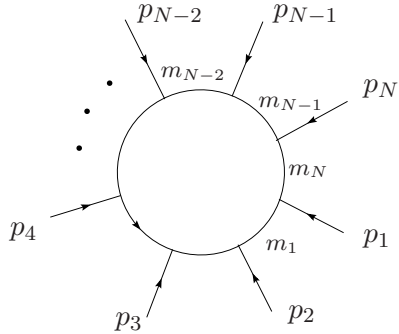}}
\put(50,16){$p_{4}$}
\put(64,4){$p_{3}$}
\put(87,5){$p_{2}$}
\put(97,15){$p_{1}$}
\put(97,35){$p_{N}$}
\put(85,45){$p_{N-1}$}
\put(67, 45){$p_{N-2}$}
\put(84,14){\footnotesize $m_{1}$}
\put(87,24){\footnotesize $m_N$}
\put(83,33){\footnotesize $m_{N-1}$}
\put(70,37){\footnotesize $m_{N-2}$}
\end{picture}
\caption{Momentum and mass assignments for a general $N$-point one-loop graph.}
\label{fig:Npoint}
\end{figure}

To be specific, for a general $N$-point integral as shown in Fig.~\ref{fig:Npoint},  
we use
\begin{eqnarray}
I^{D}_N(\{s_{j_1\ldots j_n}\};\{m_i^2\}) &=& 
\int \frac{d^D k}{i \, \pi^{D/2}}
\; \frac{\mu^{4-D}}{
[(k+r_1)^2-m_1^2+i\delta]\dots [(k+r_{N})^2-m_N^2+i\delta]}
\label{eq0}\\
&=&\mu^{4-D}(-1)^N\Gamma(N-\frac{D}{2})\int \limits_{0}^{1} \left(\prod\limits_{i=1}^N dz_i\right)\,
\frac{\delta(1-\sum_{l=1}^N z_l)}{\left(-\frac{1}{2}z_i \mathcal{S}_{ij}z_j-i\,\delta\right)^{N-\frac{D}{2}}}\;,
\nn\end{eqnarray} 
where $r_i=\sum_{j=1}^i p_j$ and $\mathcal{S}_{ij}=(r_i-r_j)^2-m_i^2-m_j^2$.
The results for the integrals are characterised by the invariants 
$s_{j_1\ldots j_n}=(p_{j_1}+\ldots+p_{j_n})^2$ and $m_i^2$, 
where the list $\{s_{j_1\ldots j_n}\}$ contains the invariants  defined by 
$n$-particle cuts of the diagram.
To map the labelling of ${\cal S}_{ij}$ resp. Fig.~\ref{fig:Npoint} to the LoopTools conventions, 
the argument lists of triangles and boxes are given by  
$I_3^D(s_1,s_2,s_3;m_3^2,m_1^2,m_2^2)$ and 
$I_4^D(s_1,s_2,s_3,s_4;s_{12},s_{23};m_4^2,m_1^2,m_2^2,m_3^2)$.


The integrals listed below will also be posted shortly to the 
{\em LoopForge} integral database at 
 {\it http://www.ippp.dur.ac.uk/LoopForge}, 
together with some numerical benchmark points.
The triangles (\ref{eq:triXs100Xm000}), (\ref{eq:triXs120Xm000}), 
(\ref{eq:triXs100Xm100}), (\ref{eq:triXs120Xm100})  and the boxes 
 (\ref{eq:boxXs1000Xm0000}), (\ref{eq:boxXs1200Xm0000}), 
 (\ref{eq:boxXs1030Xm0000}), (\ref{eq:boxXs1000Xm10004D}),
 (\ref{eq:boxXs1200Xm1100}) can also be found at 
{\it http://qcdloop.fnal.gov/} for certain kinematic regions.

For the analytical representations given below
we use the following auxiliary functions.\\
The K\"all\'en function:
\begin{equation}
  \label{eq:kaellen}
\lambda(x,y,z)=x^2+y^2+z^2-2xy-2yz-2zx,
\end{equation}
the $\eta$-function:
\begin{eqnarray}
  \label{eq:etafunc}
  \eta(x,y)&=&\log(xy)-\log(x)-\log(y), 
\end{eqnarray}
and the $R$-function \cite{'tHooft:1978xw,vanOldenborgh:1989wn}:
\begin{eqnarray}
  \label{eq:Rfunc1st}
R(y_0,z\pm \idel) &=& \int\limits_{0}^{1} \ud y \frac{ \log( y-z\mp \idel ) - \log( y_0-z\mp \idel )
}{y-y_0}\nonumber \\
&=& \textrm{Li}_2(z_1) - \textrm{Li}_2(z_2) + \eta_1 \; \log(z_1) - \eta_2 \; \log(z_2) \\
{\rm where} \hspace{1.5cm} z_1 &=& \frac{y_0}{y_0-z\mp \idel}\;, \hspace{0.5cm} z_2 =
\frac{y_0-1}{y_0-z\mp \idel} \nonumber\\ 
\eta_1 &=& \eta( -z\mp \idel,  1/(y_0-z\mp \idel) )\;, \hspace{0.5cm} 
\eta_2 = \eta( 1-z\mp \idel, 1/(y_0-z\mp \idel) ).\nonumber
\end{eqnarray}
We work in general in $D=4-2\epsilon$ dimensions but give all formulas only up to 
${\cal O}(\varepsilon^0)$. We abbreviate infinitesimal displacements in the Mandelstam 
variables as
$\tilde{s}=s+\idel$. 


\vspace*{8mm}

\subsection*{Triangle integrals}

In the figures below, a single (double) internal line represents 
a massless (massive) propagator, while a single
(double) external leg represents one for which $p^2$ is zero (non-zero).

\unitlength=8.bp
  \begin{feynartspicture}(50,10)(1,1)
    \FADiagram{}
    \FAProp(6.,5.)(10.,11.)(0.,){/Straight}{0}
    \FAProp(6.,5.)(14.,5.)(-0.,){/Straight}{0}
    \FAProp(10.,11.)(14.,5.)(-0.,){/Straight}{0}
    \FAProp(6.,5.2)(1.,3.2)(0.,){/Straight}{0}
    \FAProp(6.1,4.8)(1.1,2.8)(-0.,){/Straight}{0}
    \FAProp(19.,3.)(14.,5.)(0.,){/Straight}{0}
    \FAProp(10.,11.)(10.,15.)(-0.,){/Straight}{0}
    \put(-2,4){{\bf Triangle} $I_3^D(s_1,0,0;0,0,0)$}
    \put(20.5,0.9){$p_1$}
  \end{feynartspicture} 
\begin{equation}
  \label{eq:triXs100Xm000}
  I_3^D(s_1,0,0;0,0,0)= \frac{\Gamma(1+\varepsilon)}{s_1} \left[
    \frac{1}{\varepsilon^2}-\frac{1}{\varepsilon}\log\left(\frac{-\tilde{s}_1}{\mu^2}\right)+\half
    \log^2\left(\frac{-\tilde{s}_1}{\mu^2}\right)-\frac{\pi^2}{6} \right] \quad .
\end{equation}


\newpage

\unitlength=8.bp
  \begin{feynartspicture}(50,10)(1,1)
    \FADiagram{}
    \FAProp(6.,5.)(10.,11.)(0.,){/Straight}{0}
    \FAProp(6.,5.)(14.,5.)(-0.,){/Straight}{0}
    \FAProp(10.,11.)(14.,5.)(-0.,){/Straight}{0}
    \FAProp(19.,3.)(14.,5.)(0.,){/Straight}{0}
    \FAProp(6.,5.2)(1.,3.2)(0.,){/Straight}{0}
    \FAProp(6.1,4.8)(1.1,2.8)(-0.,){/Straight}{0}
    \FAProp(9.8,11.)(9.8,15.)(-0.,){/Straight}{0}
    \FAProp(10.2,11.)(10.2,15.)(-0.,){/Straight}{0}
   \put(-2,4){{\bf Triangle}  $I_3^D(s_1,s_2,0;0,0,0)$}
    \put(20.5,0.9){$p_1$}
    \put(23,6.5){$p_2$}
  \end{feynartspicture}
  
\begin{eqnarray}
  \label{eq:triXs120Xm000}
  I_3^D(s_1,s_2,0;0,0,0)&=&\\ 
  &&\nonumber\\
  \nonumber && \hspace{-4.7cm}
  \frac{\Gamma(1+\varepsilon)}{s_1-s_2} \left\{
        \frac{1}{\varepsilon} \left[ \log\left(\frac{-\tilde{s}_2}{\mu^2}\right) -
          \log\left(\frac{-\tilde{s}_1}{\mu^2}\right) \right] - \half
        \log^2\left(\frac{-\tilde{s}_2}{\mu^2}\right) + \half 
        \log^2\left(\frac{-\tilde{s}_1}{\mu^2}\right) \right\} .
\end{eqnarray}


\unitlength=8.bp
  \begin{feynartspicture}(50,10)(1,1)
    \FADiagram{}
    \FAProp(6.,5.)(10.,11.)(0.,){/Straight}{0}
    \FAProp(6.,5.)(14.,5.)(-0.,){/Straight}{0}
    \FAProp(10.,11.)(14.,5.)(-0.,){/Straight}{0}
    \FAProp(6.,5.2)(1.,3.2)(0.,){/Straight}{0}
    \FAProp(6.1,4.8)(1.1,2.8)(-0.,){/Straight}{0}
    \FAProp(18.9,2.8)(13.9,4.8)(0.,){/Straight}{0}
    \FAProp(19.,3.2)(14.,5.2)(0.,){/Straight}{0}
    \FAProp(9.8,11.)(9.8,15.)(-0.,){/Straight}{0}
    \FAProp(10.2,11.)(10.2,15.)(-0.,){/Straight}{0}
    \put(-2,4){{\bf Triangle}  $I_3^D(s_1,s_2,s_3;0,0,0)$}
  \end{feynartspicture} 
  
   \begin{eqnarray}
  \label{eq:triXs123Xm000}
  I_3^D(s_1,s_2,s_3;0,0,0)&=&  -\frac{1}{\sqrt{\lambda(s_1,s_2,s_3) - i\delta\;s_1}}\;
\Bigl[2{\rm Li}_2\left(-\frac{x_-}{y_+}\right) +
2{\rm Li}_2\left(-\frac{y_-}{x_+}\right) + \frac{\pi^2}{3} \nonumber \\ 
&& \hspace{-3cm} 
+\frac{1}{2} \log^2\left( \frac{x_-}{y_+} \right)
+\frac{1}{2} \log^2\left( \frac{y_-}{x_+} \right)
+\frac{1}{2} \log^2\left( \frac{x_+}{y_+} \right)
-\frac{1}{2} \log^2\left( \frac{x_-}{y_-} \right)
\Bigr] 
\end{eqnarray}
 with 
 \begin{eqnarray}
 x_\pm &=& \frac{s_1+s_3-s_2 \mp \sqrt{\lambda(s_1,s_2,s_3) - i\delta\;s_1  }}{2\;s_1}\quad ,\quad
y_\pm = 1-x_\mp.\nonumber  
\end{eqnarray}
Note that the permutation symmetry of the integral in $s_1,s_2,s_3$ is preserved, 
although not manifest due to the choice of the denominator in $x_\pm$.

\newpage



 \unitlength=8.bp
  \begin{feynartspicture}(50,10)(1,1)
    \FADiagram{}
    \FAProp(6.,5.)(10.,11.)(0.,){/Straight}{0}
    \FAProp(6.5,5.)(10.3,10.6)(0.,){/Straight}{0}
    \FAProp(6.,5.)(14.,5.)(-0.,){/Straight}{0}
    \FAProp(10.,11.)(14.,5.)(-0.,){/Straight}{0}
    \FAProp(6.,5.2)(1.,3.2)(0.,){/Straight}{0}
    \FAProp(6.1,4.8)(1.1,2.8)(-0.,){/Straight}{0}
    \FAProp(19.,3.)(14.,5.)(0.,){/Straight}{0}
    \FAProp(10.,11.)(10.,15.)(-0.,){/Straight}{0}
    \put(-2,4){{\bf Triangle} $I_3^D(s_1,0,0;0,M^2,0)$}
  \end{feynartspicture}

  \begin{eqnarray}
  \label{eq:triXs100Xm100}
  I_3^D(s_1,0,0;0,M^2,0) &=& \frac{\Gamma(1+\varepsilon)}{-s_1} \left\{
    \frac{1}{\varepsilon} \left[ \log\left( \frac{-\tilde{s}_1+M^2}{\mu^2}\right) - \log\left(
        \frac{M^2}{\mu^2}\right) \right] \right. \\ \nonumber  
& & \hspace{-4cm} \left. + {\rm Li}_2\left( \frac{\tilde{s}_1}{\tilde{s}_1-M^2 } \right) 
 - \frac{1}{2} \left[  \log^2\left( \frac{-\tilde{s}_1+M^2}{\mu^2} \right) -  \log^2\left(
   \frac{M^2}{\mu^2}\right) \right] \right\}
\end{eqnarray}


\unitlength=8.bp 
  \begin{feynartspicture}(50,10)(1,1)
    \FADiagram{}
    \FAProp(6.,5.)(10.,11.)(0.,){/Straight}{0}
    \FAProp(6.,5.)(14.,5.)(-0.,){/Straight}{0}
    \FAProp(6.5,5.)(10.3,10.6)(0.,){/Straight}{0}
    \FAProp(10.,11.)(14.,5.)(-0.,){/Straight}{0}
    \FAProp(19.,3.)(14.,5.)(0.,){/Straight}{0}
    \FAProp(6.,5.2)(1.,3.2)(0.,){/Straight}{0}
    \FAProp(6.1,4.8)(1.1,2.8)(-0.,){/Straight}{0}
    \FAProp(9.8,11.)(9.8,15.)(-0.,){/Straight}{0}
    \FAProp(10.2,11.)(10.2,15.)(-0.,){/Straight}{0}
 \put(-2,4){{\bf Triangle } $I_3^D(s_1,s_2,0;0,M^2,0)$}
  \end{feynartspicture} 

\begin{eqnarray}
  \label{eq:triXs120Xm100}
  I_3^D(s_1,s_2,0;0,M^2,0) &=& \frac{\Gamma(1+\varepsilon)}{s_2-s_1}
  \left\{\frac{1}{\varepsilon} \left[  \log\left(\frac{-\tilde{s}_1+M^2}{\mu^2}\right) -
      \log\left(\frac{-\tilde{s}_2+M^2}{\mu^2}\right) \right] \right. \nonumber \\
& &  \hspace{-2.5cm}+\textrm{Li}_2\left(
  \frac{\tilde{s}_1}{\tilde{s}_1-M^2} \right) - \textrm{Li}_2\left(
  \frac{\tilde{s}_2}{\tilde{s}_2-M^2} \right)  \nonumber  \\ 
& & \hspace{-2.5cm}\left. +  \frac{1}{2} \log^2\left(\frac{-\tilde{s}_2+M^2}{\mu^2}\right) -
  \frac{1}{2} \log^2\left(\frac{-\tilde{s}_1+M^2}{\mu^2}\right)  
 \right\}
\end{eqnarray}


  \unitlength=8.bp  
  \begin{feynartspicture}(50,10)(1,1)
    \FADiagram{}
    \FAProp(6.,5.)(10.,11.)(0.,){/Straight}{0}
    \FAProp(6.,5.)(14.,5.)(-0.,){/Straight}{0}
    \FAProp(10.,11.)(14.,5.)(-0.,){/Straight}{0}
    \FAProp(13.5,5.)(9.7,10.6)(0.,){/Straight}{0}    
    \FAProp(19.,3.)(14.,5.)(0.,){/Straight}{0}
    \FAProp(6.,5.2)(1.,3.2)(0.,){/Straight}{0}
    \FAProp(6.1,4.8)(1.1,2.8)(-0.,){/Straight}{0}
    \FAProp(9.8,11.)(9.8,15.)(-0.,){/Straight}{0}
    \FAProp(10.2,11.)(10.2,15.)(-0.,){/Straight}{0}
 \put(-2,4){{\bf Triangle } $I_3^D(s_1,s_2,0;0,0,M^2)$}
  \end{feynartspicture} 
  
\begin{eqnarray}
  \label{eq:triXs120Xm010}
  I_3^D(s_1,s_2,0;0,0,M^2) &=& \frac{1}{s_2-s_1} \left\{R(x_0,\tilde
    x_1)-\frac{\pi^2}{6}+{\rm Li}_2\left(1-\frac{1}{x_0-i\delta} \right) \right\} \\
  \nonumber  {\rm with} \hspace{3mm} x_0 &=& \frac{s_1}{s_1-s_2}\;, \;
     \tilde{x}_1 = \frac{\tilde{s}_1}{s_1-s_2+M^2}
\end{eqnarray}


 \unitlength=8.bp 
  \begin{feynartspicture}(50,10)(1,1)
    \FADiagram{}
    \FAProp(6.,5.)(10.,11.)(0.,){/Straight}{0}
    \FAProp(6.,5.)(14.,5.)(-0.,){/Straight}{0}
    \FAProp(10.,11.)(14.,5.)(-0.,){/Straight}{0} 
    \FAProp(6.,5.2)(1.,3.2)(0.,){/Straight}{0}
    \FAProp(6.1,4.8)(1.1,2.8)(-0.,){/Straight}{0}
    \FAProp(9.8,11.)(9.8,15.)(-0.,){/Straight}{0}
    \FAProp(10.2,11.)(10.2,15.)(-0.,){/Straight}{0}
    \FAProp(19.,3.)(14.,5.)(0.,){/Straight}{0}
    \FAProp(13.5,5.)(9.7,10.6)(0.,){/Straight}{0}    
    \FAProp(6.5,5.)(10.3,10.6)(0.,){/Straight}{0}
  \put(-2,4){{\bf Triangle} $ I_3^D(s_1,s_2,0;0,M^2,M^2)$} 
  \end{feynartspicture}

\begin{eqnarray}
  \label{eq:triXs120Xm110}
  I_3^D(s_1,s_2,0;0,M^2,M^2)&=& \\ \nonumber && \hspace{-4cm} \frac{1}{s_2-s_1} \left\{
    {\rm Li}_2\left( \frac{\tilde{s}_1}{M^2}  \right)-{\rm 
      Li}_2\left(\frac{1}{x_+} \right) -{\rm Li}_2\left(\frac{1}{x_-} \right)+R\left( x_0,
      \frac{M^2}{\tilde{s}_1} \right) \right. \\ \nonumber && \hspace{-3.5cm} \left. +
  R(1-x_0,x_-)-R(x_0,x_-)-\eta_0 \log\left( \frac{1-x_0}{-x_0} \right) \right\} 
\end{eqnarray}  

with
\begin{eqnarray}
x_0&=&1-\frac{s_1}{s_2} \;,\;x_\pm=\half\left( 1\pm \sqrt{1-\frac{4M^2}{
      \tilde{s}_2}}\right) \nn\\ \nonumber
\eta_0&=&\eta\left(1-\frac{\tilde{s}_1}{M^2} x_0,\frac{M^2}{M^2 - \tilde{s}_2 x_0(1-x_0)} \right)
\end{eqnarray}

\unitlength=8.bp
  \begin{feynartspicture}(50,10)(1,1)
    \FADiagram{}
    \FAProp(6.,5.)(10.,11.)(0.,){/Straight}{0}
    \FAProp(6.,5.)(14.,5.)(-0.,){/Straight}{0}
    \FAProp(10.,11.)(14.,5.)(-0.,){/Straight}{0} 
    \FAProp(6.,5.2)(1.,3.2)(0.,){/Straight}{0}
    \FAProp(6.1,4.8)(1.1,2.8)(-0.,){/Straight}{0}
    \FAProp(9.8,11.)(9.8,15.)(-0.,){/Straight}{0}
    \FAProp(10.2,11.)(10.2,15.)(-0.,){/Straight}{0}
    \FAProp(18.9,2.8)(13.9,4.8)(0.,){/Straight}{0}
    \FAProp(19.,3.2)(14.,5.2)(0.,){/Straight}{0}
    \FAProp(13.5,5.)(9.7,10.6)(0.,){/Straight}{0}    
    \FAProp(6.5,5.)(10.3,10.6)(0.,){/Straight}{0}
    \put(-2,4){{\bf Triangle}  $I_3^D(s_1,s_2,s_3;0,M^2,M^2)$}
  \end{feynartspicture}
  
  \begin{eqnarray}
  \label{eq:triXs123Xm110}
  I_3^D(s_1,s_2,s_3;0,M^2,M^2) &=& \\ \nonumber && \hspace{-4.5cm}
  \frac{1}{\sqrt{\lambda(s_1,s_2,s_3)}} \left[ 
    R(x_-,\tilde x_1)-R(x_+,\tilde x_1) 
  +R(1-x_-,\tilde x_0)-R(1-x_+,\tilde x_0) \phantom{\frac{1}{1}} \right.  \\
  \nonumber && \hspace{-4.5cm} -R(x_-,\tilde x_0)+R(x_+,\tilde
  x_0)\left. -\eta_-\log\left( \frac{1-x_-}{-x_-} \right)
    +\eta_+\log\left( \frac{1-x_+}{-x_+} \right) \right]  
   \end{eqnarray} 
    with
  \begin{eqnarray}
   x_\pm &=& \frac{s_1+s_2-s_3\mp \sqrt{\lambda(s_1,s_2,s_3)}}{2s_2} \nonumber \\
  \tilde x_0 &=& \half\left(1-\sqrt{1-\frac{4M^2}{\tilde{s}_2}}\right) \;,\;
  \tilde{x}_1 = \frac{\tilde{s}_1-M^2}{s_1-s_3} \nonumber\\ 
  \eta_\pm &=& \eta\left(-s_3 x_\pm
    -s_1(1-x_\pm)+M^2-i\delta,\phantom{\frac{1}{1}} \frac{1}{M^2-x_\pm(1-x_\pm) \tilde{s}_2 }\right)\;.\nonumber
\end{eqnarray}
In this formula we assume momentum conservation and at least one positive invariant $s_j$.
The latter condition is always guaranteed if the triangle graph is a subgraph of $1 \to
n$ or $2 \to n$ scattering kinematics. These lead to a positive Kaellen function:
$\lambda(s_1,s_2,s_3) > 0$.


\vspace*{1cm}
  
\subsection*{Box integrals}

The results are given here for $D$=4-2$\epsilon$ dimensional boxes.  
The conversion relation to 6-dimensional boxes is achieved by the formula
\beq\label{4D26D}
I_4^D=\sum_{i=1}^4 b_i\,I^D_{3,i}+(D-3)\,B\,I_4^{D+2}\;,
\eeq
where $b_i=\sum_{j=1}^4 \mathcal{S}^{-1}_{ij}$ and  $B=\sum_{i=1}^4\, b_i$. 
$I^D_{3,i}$ denotes the ``pinch"  triangle 
stemming from a box where the {\it i}th propagator is omitted. All pinch triangle integrals
 needed by eq.~(\ref{4D26D}) are given above. 
 We use $s_{12}=s$ and $s_{23}=t$ in the following.



    \unitlength=8.bp
  \begin{feynartspicture}(50,10)(1,1)
    \FADiagram{}
    \FAProp(6.,5.)(6.,13.)(0.,){/Straight}{0}
    \FAProp(6.,5.)(14.,5.)(-0.,){/Straight}{0}
    \FAProp(6.,13.)(14.,13.)(-0.,){/Straight}{0}
    \FAProp(14.,5.)(14.,13.)(-0.,){/Straight}{0}
    \FAProp(6.,5.2)(1.,3.2)(0.,){/Straight}{0}
    \FAProp(6.1,4.8)(1.1,2.8)(-0.,){/Straight}{0}
    \FAProp(6.,13.)(1.,15.)(0.,){/Straight}{0}
    \FAProp(14.,13.)(19.,15.)(0.,){/Straight}{0}
    \FAProp(14.,5.)(19.,3.)(0.,){/Straight}{0}
    \put(-2,4){{\bf Box} $I_4^D(s_1,0,0,0;s,t;0,0,0,0)$}
  \end{feynartspicture} 
  
\begin{eqnarray}
  \label{eq:boxXs1000Xm0000}
  I_4^D(s_1,0,0,0;s,t;0,0,0,0) &=& \frac{\Gamma(1+\varepsilon)}{st}
  \left\{ \frac{2}{\varepsilon^2}-\frac{2}{\epsilon}
  \left[\log\left(\frac{-\tilde{s}}{\mu^2} \right)
       +\log\left(\frac{-\tilde{t}}{\mu^2} \right)
       -\log\left(\frac{-\tilde{s}_1}{\mu^2}\right)\right]
  \right.  \nonumber \\  && \hspace{-3cm}   
  +2\,{\rm Li}_2\left(1-\frac{\tilde{s}}{\tilde{s}_1}\right) 
  +2\,{\rm Li}_2\left(1-\frac{\tilde{t}}{\tilde{s}_1}\right) 
  +2\log \left(
      \frac{\tilde{s}}{\tilde{s}_1} \right)\log \left(
      \frac{\tilde{t}}{\tilde{s}_1} \right)\\ && \hspace{-3cm}
    \left. 
      +\log^2\left(\frac{-\tilde{s}}{\mu^2}\right)
      -\log^2\left(\frac{-\tilde{s}_1}{\mu^2}\right)
      +\log^2\left(\frac{-\tilde{t}}{\mu^2}\right)-\frac{2\pi^2}{3}
    \phantom{\frac{1}{1}} \hspace{-0.3cm} \right\}\nn
\end{eqnarray}




 \begin{feynartspicture}(50,10)(1,1)
    \FADiagram{}
    \FAProp(6.,5.)(6.,13.)(0.,){/Straight}{0}
    \FAProp(6.,5.)(14.,5.)(-0.,){/Straight}{0}
    \FAProp(6.,13.)(14.,13.)(-0.,){/Straight}{0}
    \FAProp(14.,5.)(14.,13.)(-0.,){/Straight}{0}
    \FAProp(6.,5.2)(1.,3.2)(0.,){/Straight}{0}
    \FAProp(6.1,4.8)(1.1,2.8)(-0.,){/Straight}{0}
    \FAProp(6.,12.8)(1.,14.8)(0.,){/Straight}{0}
    \FAProp(6.1,13.2)(1.1,15.2)(0.,){/Straight}{0}
    \FAProp(14.,13.)(19.,15.)(0.,){/Straight}{0}
    \FAProp(14.,5.)(19.,3.)(0.,){/Straight}{0}
    \put(-2,4){{\bf Box} $I_4^D(s_1,s_2,0,0;s,t;0,0,0,0)$}
  \end{feynartspicture} 
    
\begin{eqnarray}
  \label{eq:boxXs1200Xm0000}
  I_4^D(s_1,s_2,0,0;s,t;0,0,0,0) &=& \\ \nonumber && \hspace{-5cm}
  \frac{\Gamma(1+\varepsilon)}{st} 
  \left\{\frac{1}{\varepsilon^2} +
    \frac{1}{\varepsilon}\left[\log\left(\frac{-\tilde{s}_1}{\mu^2}\right) +
      \log\left(\frac{-\tilde{s}_2}{\mu^2}\right) -
      \log\left(\frac{-\tilde{s}}{\mu^2}\right)-2\log\left(\frac{-\tilde{t}}{\mu^2}\right)\right]
  -\frac{\pi^2}{6}  \right. \\ \nonumber && \hspace{-5cm} -2{\rm
      Li}_2\left(1-\frac{\tilde{s}_1}{\tilde{t}}\right)-2{\rm
      Li}_2\left(1-\frac{\tilde{s}_2}{\tilde{t}}\right) +
      \log\left(\frac{\tilde{s}}{\tilde{s}_1}\right)
      \log\left(\frac{\tilde{s}}{\tilde{s}_2}\right) \\
      \nonumber && \hspace{-5cm} 
      \left.-\log^2\left(\frac{\tilde{s}}{\tilde{t}}\right) +
        \half\log^2\left(\frac{-\tilde{s}}{\mu^2}\right) - 
        \half\log^2\left(\frac{-\tilde{s}_1}{\mu^2}\right) - 
        \half\log^2\left(\frac{-\tilde{s}_2}{\mu^2}\right) + \log^2\left(\frac{-\tilde{t}}{\mu^2}\right)
        \phantom{\frac{1}{1}} \hspace{-0.4cm}\right\}
\end{eqnarray}


 \begin{feynartspicture}(50,10)(1,1)
    \FADiagram{}
    \FAProp(6.,5.)(6.,13.)(0.,){/Straight}{0}
    \FAProp(6.,5.)(14.,5.)(-0.,){/Straight}{0}
    \FAProp(6.,13.)(14.,13.)(-0.,){/Straight}{0}
    \FAProp(14.,5.)(14.,13.)(-0.,){/Straight}{0}
    \FAProp(6.,5.2)(1.,3.2)(0.,){/Straight}{0}
    \FAProp(6.1,4.8)(1.1,2.8)(-0.,){/Straight}{0}
    \FAProp(6.,13.)(1.,15.)(0.,){/Straight}{0}
    \FAProp(14.,12.8)(19.,14.8)(0.,){/Straight}{0}
    \FAProp(13.9,13.2)(18.9,15.2)(0.,){/Straight}{0}
    \FAProp(14.,5.)(19.,3.)(0.,){/Straight}{0}
    \put(-2,4){{\bf Box} $I_4^D(s_1,0,s_3,0;s,t;0,0,0,0)$}
  \end{feynartspicture} 

\begin{eqnarray}
  \label{eq:boxXs1030Xm0000}
  I_4^D(s_1,0,s_3,0;s,t;0,0,0,0)&=& \\ \nonumber &&
  \hspace{-5cm}\frac{\Gamma(1+\varepsilon)}{st-s_1s_3} \left\{
    \frac{2}{\epsilon}
    \left[\log\left(\frac{-\tilde{s}_1}{\mu^2}\right)
         +\log\left(\frac{-\tilde{s}_3}{\mu^2}\right)
	 -\log\left(\frac{-\tilde{s}}{\mu^2}\right)
	 -\log\left(\frac{-\tilde{t}}{\mu^2}\right)
    \right] \right. \nn\\  
    && \hspace{-5cm}
    + \log^2\left(\frac{-\tilde{s} }{\mu^2}\right) 
    - \log^2\left(\frac{-\tilde{s}_1}{\mu^2}\right) 
    - \log^2\left(\frac{-\tilde{s}_3}{\mu^2}\right)      
    + \log^2\left(\frac{-\tilde{t} }{\mu^2}\right) \nonumber\\
  && \hspace{-5cm}  -2\,{\rm Li}_2\left(1 -\frac{\tilde{s}_1}{\tilde{s}}\right) 
     -2\,{\rm Li}_2\left(1 -
      \frac{\tilde{s}_3}{\tilde{s}}\right)   -2 \, {\rm Li}_2\left(1 -
      \frac{\tilde{s}_1}{\tilde{t}}\right) -2 \, {\rm Li}_2\left(1 -
      \frac{\tilde{s}_3}{\tilde{t}}\right) \nonumber\\  && 
      \hspace{-5cm}\left. +2 \,{\rm
      Li}_2\left(1 - \frac{\tilde{s}_1\,\tilde{s}_3}{\tilde{s}\,\tilde{t}}\right) -
     \log^2\left(\frac{\tilde{s}}{\tilde{t}}\right)  +2 \,
    \eta\left(\frac{\tilde{s}_3}{\tilde{s}},\frac{\tilde{s}_1}{\tilde{t}}\right)
    \log\left(1-\frac{\tilde{s}_1 \, \tilde{s}_3}{ \tilde{s} \,\tilde{t}}\right)
      \phantom{\frac{1}{1}} \hspace{-0.3cm} \right\}\nonumber
\end{eqnarray}

\newpage



\begin{feynartspicture}(50,10)(1,1)
    \FADiagram{}
    \FAProp(6.,5.)(6.,13.)(0.,){/Straight}{0}
    \FAProp(6.,5.)(14.,5.)(-0.,){/Straight}{0}
    \FAProp(6.,13.)(14.,13.)(-0.,){/Straight}{0}
    \FAProp(14.,5.)(14.,13.)(-0.,){/Straight}{0}
    \FAProp(6.4,5.)(6.4,13.)(0.,){/Straight}{0}
    \FAProp(6.,5.2)(1.,3.2)(0.,){/Straight}{0}
    \FAProp(6.1,4.8)(1.1,2.8)(-0.,){/Straight}{0}
    \FAProp(6.,13.)(1.,15.)(0.,){/Straight}{0}
    \FAProp(14.,13.)(19.,15.)(0.,){/Straight}{0}
    \FAProp(14.,5.)(19.,3.)(0.,){/Straight}{0}
    \put(-2,4){{\bf Box} $I_4^{D}(s_1,0,0,0;s,t;0,M^2,0,0)$}
  \end{feynartspicture} 
  
  We find for the 6-dimensional integral
  
  \begin{eqnarray}
  \label{eq:boxXs1000Xm1000}
  I_4^{D+2}(s_1,0,0,0;s,t;0,M^2,0,0) &=& \\ \nonumber && \hspace{-4cm}
  \frac{-t+M^2}{t(s_1-s-t)} \left[R\left(x_0,\tilde{x}_2\right)+ {\rm Li}_2 
    \left(1-\frac{1}{\tilde{x}_0}\right)  - \frac{\pi^2}{6}\right] \\ \nonumber &&
  \hspace{-4cm} -\frac{M^2}{t(s_1-s)}
  \left[ R\left(x_1,\tilde{x}_2\right)  + {\rm Li}_2 
    \left(1-\frac{1}{\tilde{x}_1}\right) - \frac{\pi^2}{6}\right]  \nonumber
    \end{eqnarray}
    where
    \begin{eqnarray}
  x_0= \frac{s}{s+t-s_1} &,& \tilde{x}_0=  x_0 + \frac{\idel}{t-M^2}  \; , \nonumber\\
  x_1= \frac{s}{s-s_1} &,& \tilde{x}_1= x_1 - \idel\; \;,\; \;  \tilde{x}_2= \frac{\tilde{s}}{s-s_1+M^2}\nonumber
\end{eqnarray}

The integral in $D=4-2\epsilon$ is obtained by

\begin{eqnarray}
  \label{eq:boxXs1000Xm10004D}
  I_4^{D}(s_1,0,0,0;s,t;0,M^2,0,0) &=& 
  \quad b_1\,I_3(s,0,0;0,0,0)+b_2\,I_3(s_1,t,0;0,M^2,0)\\
  &&\hspace{-5cm}
  +b_3\,I_3(s_1,0,s;0,M^2,0)+b_4\,I_3(t,0,0;0,M^2,0)+B\,I_4^{D+2}(s_1,0,0,0;s,t;0,M^2,0,0) 
  \nonumber
  \end{eqnarray}
  with 
  \begin{eqnarray}
  b_1&=&-\frac{1}{M^2 - t} \;,\;
  b_2=\frac{s_1-t}{s\,(M^2-t)}\;,\;
  b_3=\frac{t(s-s_1)+ M^2(s+2t-s_1)}{s\,(M^2 -t)^2}\nonumber\\
  b_4&=&-\frac{t}{s\,(M^2 -t)}\;,\;
  B = \frac{2\,t\,(s+t-s_1)}{s\,(M^2 -t)^2}  \;.\nn
\end{eqnarray}

\newpage



 \begin{feynartspicture}(50,10)(1,1)
    \FADiagram{}
    \FAProp(6.,5.)(6.,13.)(0.,){/Straight}{0}
    \FAProp(6.,5.)(14.,5.)(-0.,){/Straight}{0}
    \FAProp(6.,13.)(14.,13.)(-0.,){/Straight}{0}
    \FAProp(14.,5.)(14.,13.)(-0.,){/Straight}{0}
    \FAProp(6.4,5.)(6.4,13.)(0.,){/Straight}{0}
    \FAProp(6.,12.6)(14.,12.6)(0.,){/Straight}{0}
    \FAProp(6.,5.2)(1.,3.2)(0.,){/Straight}{0}
    \FAProp(6.1,4.8)(1.1,2.8)(-0.,){/Straight}{0}
    \FAProp(6.,12.8)(1.,14.8)(0.,){/Straight}{0}
    \FAProp(6.1,13.2)(1.1,15.2)(0.,){/Straight}{0}
    \FAProp(14.,13.)(19.,15.)(0.,){/Straight}{0}
    \FAProp(14.,5.)(19.,3.)(0.,){/Straight}{0}
    \put(-2,4){{\bf Box} $I_4^{D}(s_1,s_2,0,0;s,t;0,M^2,M^2,0)$}
  \end{feynartspicture} 
  
  
  \begin{eqnarray}
  \label{eq:boxXs1200Xm1100}
  I_4^{D}(s_1,s_2,0,0;s,t;0,M^2,M^2,0) &=& 
  \frac{I_{\textrm{pole}} + I_{\textrm{finite}}}{st - (s+t-s_1)M^2}
  \end{eqnarray}
 with  
  \begin{eqnarray}
  I_{\textrm{pole}} &=& -\frac{\Gamma(1+\epsilon)}{\epsilon} 
\Bigl[ 
  \log\left( \frac{-\tilde{t}+M^2}{M^2} \right)
- \log\left(\frac{-\tilde{s}_1+M^2}{-\tilde{s}+M^2} \right)
\Bigr] \nonumber\\
I_{\textrm{finite}} &=&  
  \log^2\left( \frac{-\tilde{t}+M^2}{\mu^2}\right) 
- \log^2\left( \frac{M^2}{\mu^2}\right)  
 -\log^2\left( \frac{-\tilde{s}_1+M^2}{\mu^2}\right) 
 +\log^2\left( \frac{-\tilde{s}+M^2}{\mu^2}\right)
\nn\\&& - 2\,R(x_0,\tilde{x}_1) + 2\, R(x_0,\tilde{x}_2)\nn\\
 && +R(1-x_1,x_-) - R(x_1,x_-) - R(1-x_2,x_-) + R(x_2,x_-) \nn\\
     && - \Bigl[ \log\left( -\tilde{s}_2/\mu^2 \right) + \log( x_+ -x_1 ) + \log( x_1 - x_- )
     \Bigr] \log\left(\frac{1-\tilde{x}_1}{-\tilde{x}_1}\right) \nn\\
     && + \Bigl[ \log\left( -\tilde{s}_2/\mu^2  \right) + \log( x_+ -x_2 ) + \log( x_2 - x_- )
     \Bigr]\log\left(\frac{1-\tilde{x}_2}{-\tilde{x}_2}\right)\nn\\ 
 x_{\pm} &=& \frac{1}{2}\left( 1 \pm \sqrt{ 1-\frac{4 M^2}{\tilde{s}_2}} \right) \quad , 
 \quad x_0 = \frac{s}{s+t-s_1}\nn\\ 
x_1 &=&\frac{M^2}{t} \quad , \quad
\tilde{x}_1 =\frac{M^2-\idel}{t} \quad , \quad 
x_2 =\frac{s-M^2}{s-s_1} \quad , \quad 
\tilde{x}_2 =\frac{\tilde{s}-M^2}{s-s_1} \nn
\end{eqnarray}

%
  

 \begin{feynartspicture}(50,10)(1,1)
    \FADiagram{}
    \FAProp(6.,5.)(6.,13.)(0.,){/Straight}{0}
    \FAProp(6.,5.)(14.,5.)(-0.,){/Straight}{0}
    \FAProp(6.,13.)(14.,13.)(-0.,){/Straight}{0}
    \FAProp(14.,5.)(14.,13.)(-0.,){/Straight}{0}
    \FAProp(6.,12.6)(14.,12.6)(0.,){/Straight}{0}
    \FAProp(13.6,5.)(13.6,13.)(0.,){/Straight}{0}
    \FAProp(6.,5.2)(1.,3.2)(0.,){/Straight}{0}
    \FAProp(6.1,4.8)(1.1,2.8)(-0.,){/Straight}{0}
    \FAProp(6.,13.)(1.,15.)(0.,){/Straight}{0}
    \FAProp(14.,12.8)(19.,14.8)(0.,){/Straight}{0}
    \FAProp(13.9,13.2)(18.9,15.2)(0.,){/Straight}{0}
    \FAProp(14.,5.)(19.,3.)(0.,){/Straight}{0}
    \put(-2,4){{\bf Box } $I_4^D(s_1,0,s_3,0;s,t;0,0,M^2,M^2)$:}
  \end{feynartspicture} 

\begin{eqnarray}
  \label{eq:boxXs1030Xm0110}
  I_4^D(s_1,0,s_3,0;s,t;0,0,M^2,M^2) &=& \frac{I_- - I_+}{\sqrt{ \det(\mathcal{S})-Ji\delta}} 
 \end{eqnarray} 
  with the auxiliary integrals\footnote{The $\pm$ label of the integrals $I_\pm$ are 
  related to $x_\pm$. The exchange of $\pm$ has no effect on $x_0^-$ and $x_0^+$.}
 \begin{eqnarray} 
 I_\pm &=& R\left(x_\pm,\frac{M^2}{\tilde{t}}\right) - R\left(1-x_\pm,\frac{M^2}{\tilde{s}}\right)
        +R( 1-x_\pm, x_0^-)-R( x_\pm, x_0^-)   \nonumber\\ &&
	+\log\left(\frac{1-x_\pm}{-x_\pm}\right) 
	\Bigl[ 
	  \log\left(\frac{\tilde{s}}{\tilde{s}_1} \right) 
	+ \log\left(\frac{\tilde{t}}{\tilde{s}_3}  \right) + \log\left(x_\pm - \frac{M^2}{ \tilde{t}} \right)\nonumber\\
&& \qquad - \log( x_0^+ - x_\pm  )-\log(x_\pm -x_0^-) 
         +\log\left( 1-x_\pm - \frac{M^2}{ \tilde{s} }\right)\Bigr]
	\end{eqnarray}  
	where
	\begin{eqnarray}
      \det(\mathcal{S})&=&[st-s_1 s_3+M^2(s-t)]^2+4M^2 (st-s_1s_3)(-s+s_1+M^2) \nonumber\\ \nonumber
        J&=&2(s_3-t)[st-s_1s_3+M^2(s-t)]+4(st-s_1s_3)(-s+s_1+M^2), \\ \nonumber
        x_\pm&=&\frac{(st-s_1s_3)+M^2(s-t)\pm \sqrt{\det(\mathcal{S})-Ji\delta}}{2\,(st-s_1s_3)}\;\;,\;\; 
        x^\pm_0=\half\left(1\pm \sqrt{1-\frac{4M^2}{\tilde{s}_3}} \right) \nonumber
\end{eqnarray}

\subsection*{Checks on the integrals}

We tested  the Master Integrals 
by comparing our results numerically with \texttt{LoopTools} \cite{Hahn:1998yk}.  For the
IR finite box and triangle integrals this is straightforward. The IR divergent box
integrals have been checked indirectly by mapping to the 6D case which is IR finite. Using
eq. (\ref{4D26D}) the $1/\varepsilon$ poles cancel when combining the $D$-dimensional box
with the triangle pinch integrals.  The same expression can be evaluated with
\texttt{LoopTools} by using a mass regulators in the IR divergent integrals. In the given
IR finite combinations the cut-off dependence is polynomial and can be made arbitrarily
small.  We have tested the formulae in all kinematically distinguishable regions.  For
example, $I_3^{D}(s_1,s_2,s_3;0,M^2,M^2)$ has been checked in the regions resulting from
combining the conditions ($s_1<0$, $0<s_1<M^2$, $M^2<s_1$), ($s_2<0$, $0<s_2<4\,M^2$,
$4\,M^2<s_2$) and ($s_3<0$, $0<s_3<M^2$, $M^2<s_3$) such that the Kaellen function is
positive in line with the comment below equation (\ref{eq:triXs123Xm110}).


\end{appendix}

\bibliographystyle{JHEP}
\providecommand{\href}[2]{#2}\begingroup\raggedright\endgroup

\end{document}